\documentclass[usenatbib]{aa}
\include{epsf}
\usepackage{graphicx}
\usepackage{epstopdf}
\usepackage{bm}
\def\apjl{{ApJ}}		
\def\apjs{{ApJS}}		
\def\ao{{Appl.~Opt.}}		
\def\aap{{A\&A}}		

\usepackage{color}
\usepackage{amsmath}	
\usepackage{amssymb}

\def \farcs{\hbox{$.\!\!^{\prime\prime}$}}

\def \Euclid{\hbox{\it Euclid}}
\def \sextractor{\hbox{\sc SExtractor}}
\def \sersic{\hbox{S{\'e}rsic}}
\def \metacal{\hbox{\sc MetaCalibration}}
\def \metadet{\hbox{\sc MetaDetection}}
\def \galsim{\hbox{\sc GalSim}}

\defcitealias{Hoekstra17}{H17}
\defcitealias{Sheldon17}{SH17}

\newcommand{\henk}[1]{{\textcolor{black}{#1}}}
\begin{document}

\title{Accounting for object detection bias in weak gravitational lensing studies}
\author{Henk Hoekstra\inst{1}\thanks{E-mail: hoekstra@strw.leidenuniv.nl}
\and
Arun Kannawadi\inst{2,1}
\and
Thomas D. Kitching\inst{3}}
\institute{Leiden Observatory, Leiden University, PO Box 9513, 2300 RA, Leiden, the  Netherlands\\
\and
Department of Astrophysical Sciences, Princeton University, 4 Ivy Lane, Princeton, NJ 08544, USA\\
\and
Mullard Space Science Laboratory, University College London,
Holmbury St Mary, Dorking, Surrey RH5 6NT, UK}


\abstract{
Weak lensing by large-scale structure is a powerful probe of cosmology if the apparent alignments in the shapes of distant galaxies can be accurately measured. Most studies have therefore focused on improving the fidelity of the shape measurements themselves, but the preceding step of object detection has been largely ignored. In this paper we study the impact of object detection for a \Euclid-like survey and show that it leads to biases that exceed requirements for the next generation of cosmic shear surveys. In realistic scenarios, blending of galaxies is \henk{an important} source of detection bias. We find that \metadet\ is able to account for blending, leading to average multiplicative biases that meet requirements for Stage IV surveys, provided a sufficiently accurate model for the point spread function is available. Further work is needed to estimate the performance for actual surveys. Combined with sufficiently realistic image simulations, this provides a viable way forward towards accurate shear estimates for Stage IV surveys.}

\keywords{cosmology: observations -- gravitational lensing}

\titlerunning{Accounting for detection bias}
\authorrunning{Hoekstra et al.}

\maketitle

\section{Introduction}

In the past decades the theoretical framework that describes the formation of cosmic structures has been tested by ever more precise observations \citep[see e.g.][for a comprehensive comparison of results]{Planck15}. Although there is discussion about small differences between cosmological parameter estimates \citep[e.g.][]{Riess19,Joudaki20}, the general agreement is remarkable given the difficulties in obtaining these results. Importantly, the main ingredients of this `concordance model' are not understood at all: dark matter and dark energy make up the bulk of the mass-energy content of the Universe, with a `mere frosting' of baryonic matter. Although a cosmological constant is an excellent fit to the current data, its unnaturally small value is by no means satisfactory. Consequently, many alternative explanations have been suggested, including modifications of the theory of general relativity \citep[see e.g.][for an overview]{Amendola18}. To distinguish between such a multitude of ideas, dramatically better observational constraints are needed.

Of particular interest is the study of the distribution of matter as a function of redshift, because it is sensitive to the growth of structure, modified gravity and the expansion history. The practical complication that most of the matter is made up of dark matter can be overcome by measuring the correlations in the ellipticities of distant galaxies that are the result of the differential deflection of light rays by intervening structures, a phenomenon called gravitational lensing. In the case that  only single images of distant galaxies are distorted by the gravitational lensing  effect, this is known as weak lensing. The amplitude of the distortion provides us with a direct measurement of the gravitational tidal field, which in turn can be used to `map' the distribution of matter directly. This makes weak lensing by large-scale structure, or cosmic shear, one of the most powerful probes to study dark energy and the growth of structure: the statistical properties of the matter distribution can be determined as a function of cosmic time. These measurements can be compared to models of structure formation, which depend on the cosmological parameters \citep[see e.g.][for a recent review]{Kilbinger15}.

The typical change in the observed ellipticity of a distant galaxy caused by gravitational lensing (known as shear) is about a percent, i.e. much smaller than the intrinsic ellipticities of galaxies. This source of statistical uncertainty can be overcome by averaging over large numbers of galaxies, although intrinsic alignments complicate this simple picture \citep[see e.g.][for reviews]{Joachimi15,Troxel15}. The cosmological lensing signal has now been measured using ground-based observations of relatively modest areas of sky \citep[see e.g.][for some recent results from stage III surveys]{Troxel18, Hildebrandt20, Hamana20} but future surveys will cover much larger fractions of the extragalactic sky, increasing the source samples accordingly.

The change in ellipticity is also smaller than the typical biases introduced by instrumental effects. Consequently, averaging the shape measurements of large ensembles of galaxies is only meaningful if these sources of bias can be corrected for to a level that renders them sub-dominant to the statistical uncertainties afforded by the survey \citep[see][for a detailed review on weak lensing systematics]{Mandelbaum18}. This will be particularly challenging for the next generation of surveys (stage IV), such as the ones carried out by \Euclid\footnote{\url{http://euclid-ec.org}}~\citep{Laureijs11} and the {\it Nancy Grace Roman} Space Telescope\footnote{\url{https://www.stsci.edu/roman}}~\citep{Spergel15} from space, and the Legacy Survey of Space and Time
by the Rubin Observatory\footnote{\url{https://www.lsst.org}} \citep{LSST09} from the ground. 

The Point Spread Function (PSF) is the dominant source of bias in the measurements of galaxy shapes, driving the desire for space-based observations \citep{Paulin-Henriksson08,Massey13}. Another complication is the fact the shapes are measured from noisy images, which can lead to biases in the ellipticity \citep[e.g.][]{Melchior12,Refregier12,Miller13,Viola14}. Given a survey design, our current understanding of these biases, and our ability to correct for them, requirements can be placed on the instrument performance, but also on the accuracy of the shape measurement algorithm. For instance, \cite{Cropper13} present a detailed breakdown for \Euclid,
which forms the basis for some of the numbers used in this paper.

Fortunately the impact of the various sources of bias can be studied by applying the shape measurement algorithm to simulated data, where the galaxy images are sheared by a known amount. Comparison with the recovered values provides an estimate of the biases. For example, \cite{Erben01} and \cite{Hoekstra02} used simulated images to examine the performance of the {\sc KSB} algorithm developed by \cite{KSB95}. Comparing a range of methods, the Shear TEsting Programme \citep[STEP;][]{STEP1,STEP2} demonstrated the importance of how a method is actually implemented. To examine the origin of the variation in performance further, the GRavitational lEnsing Accuracy Testing (GREAT) challenges \citep{GREAT08,GREAT10,GREAT3} used idealised simulations to demonstrate the importance of noise on the performance.

However, as recently shown by \citet{Hoekstra15} the actual performance of the algorithms depends crucially on the input of the simulations, such as the distribution of galaxy ellipticities and the inclusion of faint galaxies. This was studied in more detail in \citet[][H17 hereafter]{Hoekstra17} for a \Euclid-like survey. These studies showed that the fidelity of the image simulations is crucial for an accurate estimate of the overall shear bias, which depends on the bias in the shape measurements {\it and} the selection of galaxies.  \citetalias{Hoekstra17} did not consider both contributions separately, but recent studies  \citep[e.g.][]{FenechConti16, Kannawadi19} have shown that biases are already introduced in the first step of the analysis: the detection of objects. This source of bias has been largely ignored until  \cite{FenechConti16} showed that it can be as important as the shape measurement bias in ground-based surveys. More recenty, \cite{Hernandez20} showed that detection bias is also relevant for lensing studies using {\it Hubble} Space Telescope data.

Consequently, even if the shapes of the detected galaxies are somehow measured perfectly, the shear will be biased. Such a detection bias is expected because the significance with a galaxy is detected typically depends on its orientation with respect to the shear \citep{Hirata03}
or the PSF \citep{Kaiser00, Bernstein02}. In this paper we study detection bias using image simulations, similar to those used in \citetalias{Hoekstra17}. We explore how well the bias can be quantified and which parameters are most relevant. We find that blending of galaxies is the dominant source of detection bias. Such blends are absent from studies that measure shear biases using isolated galaxies (or when placed on a grid). 

To reduce shape noise, studies typically use pairs of simulated galaxies  where a second galaxy is rotated by 90 degrees (or quartets, rotated by 45 degrees). However, if one then requires that both galaxies are detected, as in \cite{Pujol19}, the detection bias is also removed. Although this is a viable approach to reduce the number of simulated images to quantify the bias introduced by the shape measurement algorithm, it is important to realise that the resulting bias cannot be applied to the actual data, but needs to be adjusted to account for detection bias. 

A further complication arises from the fact that it may not be possible to determine the shape for every detected galaxy. Hence the shape measurement step introduces additional selections, as does assigning weights to capture the fidelity of the shape measurement. Finally, to improve constraints on cosmological parameters, the source samples are split into multiple tomographic bins, using photometric redshifts. The reliance on reliable multi-band photometry introduces further selections. Those {\it selection} biases will depend on both the shape measurement algorithm and the way samples are selected. 

The setup we use in this paper is very similar to the one used in \citetalias{Hoekstra17}, and in Sect.~\ref{sec:simulations} we briefly describe the simulation setup, highlighting some of the changes we implemented. We study detection bias and its dependence of the \sextractor\ setup and the PSF in Sect.~\ref{sec:detection}. Similar to \citetalias{Hoekstra17} we explore the sensitivity to changes in the simulation input in Sect.~\ref{sec:realism}. In Sect.~\ref{sec:metacal} we quantify the performance of \metacal~\citep{Huff17,Sheldon17} as a way to avoid image simulations for the calibration of the shape measurement step. We also examine the usefulness of its extension, the so-called \metadet\ approach \citep{Sheldon19}, which aims to avoid selection biases altogether in Sect.~\ref{sec:meta_detect}. We discuss the implications of our results for future surveys in Sect.~\ref{sec:discussion}.

\section{Simulation setup}\label{sec:simulations}

The simulated images are created using the publicly available software package \galsim\footnote{\url{https://github.com/GalSim-developers/GalSim}} \citep{Rowe15}. This suite of routines was originally developed for GREAT3 \citep{Mandelbaum14,GREAT3}, but has become the \emph{de facto} standard for image simulations in the weak lensing community. As was done in \citetalias{Hoekstra17} the galaxies are described by \sersic\ profiles, with half-light radii, apparent magnitudes and \sersic\ indices $n$ drawn from a catalogue of morphological parameters measured from resolved $F606W$ images from the GEMS survey \citep{Rix04}. We only consider galaxies fainter than magnitude $m=20$ and use the morphological parameters from the GEMS catalogue for galaxies down to $m=25.4$, and normalise the counts to 36 galaxies arcmin$^{-2}$ with $20<m<24.5$. 

As shown in \citetalias{Hoekstra17} it is important to include galaxies down to $m_{\rm lim}\approx 29$, and we follow the same procedure, except that we use a flatter count slope at fainter magnitudes: we adopt a power law slope of $\alpha_{\rm faint}=0.24$ (instead of $\alpha_{\rm faint}=0.36$ using by \citetalias{Hoekstra17}), which matches the observed counts better. The intrinsic ellipticities are drawn from a Rayleigh distribution with scale parameter $\epsilon_0=0.25$, so that the mean source ellipticity is $\langle|\epsilon^{\rm s}|\rangle\approx 0.31$. We assume that the intrinsic ellipticities ${\epsilon}^{\rm s}$ do not correlate with the morphological parameters, but note that \citet{Kannawadi19} have shown that this is not the case in reality. We refer the interested reader to \citetalias{Hoekstra17} for more details on the input catalogue.

In our baseline simulations we place galaxies randomly, but with random sub-pixel offsets.  We create pairs of images, where the galaxies are placed at the same location, but rotated by 90 degrees in the rotated case. We apply the same shear to all the galaxies in such a pair by changing the true (simulated) ellipticity using \citep{Seitz97}:
\begin{equation}
  {\epsilon}^{\rm obs}=\frac{{\epsilon}^{\rm s}+{\gamma}}{1+{\gamma}^*{\epsilon}^{\rm s}},\label{eq:ellipticity}
\end{equation}
where ${\epsilon}^{\rm s}$ is the intrinsic complex ellipticity, $\gamma$ is the complex shear that is applied\footnote{The actual observable is the reduced shear $g\equiv \gamma/(1-\kappa)$, where $\kappa$ is the convergence, and $g$ should be used in Eq.~(\ref{eq:ellipticity}). However, we only consider the shear in this paper, i.e. we assume $g=\gamma$ throughout.}, and  the asterisk indicates the complex conjugate. If both galaxies of a pair are averaged, $\langle{\epsilon}^{\rm s}\rangle={0}$ and the observed ellipticity is an unbiased estimate of the shear. Hence, a non-zero detection bias implies that one of the two galaxies in a pair is not detected in a shear-dependent fashion. 

In our baseline setup, the galaxies are placed at random positions, which ignores the impact of clustering. This was studied in more detail in \citet{Martinet19} who found that faint satellite galaxies that cluster around their host galaxy do affect the bias estimates. Moreover applying a shear to a particular configuration of galaxies also changes their positions. We ignore this in our baseline simulations, but we find that shearing the positions as well as the galaxy images barely changes the results (see Sect.~\ref{sec:estimates} and Table~\ref{tab:bias} for more details). Finally, we also created images where the galaxies are placed on a grid, so that they are about 9$''$ apart, thus eliminating any blending. This provides a useful reference to compare our baseline results against.

To allow for a more direct comparison to the results presented in \citetalias{Hoekstra17}, unless specified otherwise, we use the same setup for the telescope parameters, i.e. we use a circular Airy PSF for a telescope with a diameter of 1.2m and an obscuration of 0.3 at a reference wavelength of 800nm, which is a reasonable approximation to the \Euclid\ PSF in the VIS-band \citep{Cropper18}. The individual images are 4000 pixels on a side, with a pixel size of $0\farcs1$ per pixel. The noise level is the same as used in \citetalias{Hoekstra17}, corresponding to a surface brightness of 27.7 magnitudes arcsecond$^{-2}$. This mimics the depth of 4 coadded exposures, and yields a typical number density of 47 galaxies arcmin$^{-2}$ with a signal-to-noise ratio larger than 10, as measured by \sextractor, and a number density of 33 galaxies arcmin$^{-2}$ if we restrict the magnitude range to $20<m< 24.5$.  

\begin{table}
\begin{center}
  \caption{Relevant \sextractor\ setup parameters\label{tab:setup}}
  \centering
\begin{tabular*}{\columnwidth}{ll}
  \hline
  {\tt DETECT\_MINAREA}  & 6 \\
  {\tt DETECT\_THRESH}   & 1.5\\		
  {\tt FILTER\_NAME}     & {\tt gauss\_3.0\_7x7.conv}\\
  {\tt DEBLEND\_NTHRESH} & 32\\
  {\tt DEBLEND\_MINCONT} & 0.005\\
  {\tt CLEAN\_PARAM} & 1.0\\
  {\tt BACK\_SIZE} &	600\\
  {\tt BACK\_FILTERSIZE} & 3\\
  \hline
\end{tabular*}
\bigskip
\begin{minipage}{\linewidth}
\tablefoot{Column~1: keyword in \sextractor\ configuration file;
    Column~2: value of the parameter.}
\end{minipage}
\end{center}
\end{table}

\subsection{Analysis setup}
\label{sec:setup}

We use \sextractor\ \citep{Bertin96} to detect objects in the simulated images. Our baseline setup uses the (relevant) parameter values listed in Table~\ref{tab:setup}, which are fairly standard. To detect an object, at least {\tt DETECT\_MINAREA} adjacent pixels  need to be above the threshold, which is specified by {\tt DETECT\_THRESH} times the noise level. We let \sextractor\ determine the background level, although we could have specified a global value of zero. We explored various background determination settings, and found that they did not change our results. We discuss the purpose of some of these parameters and their impact on detection bias in more detail in Sect.~\ref{sec:var_setup} and Appendix~\ref{app:setup}.

For reference, we also repeat the shape measurements using the KSB algorithm employed in \citetalias{Hoekstra17}, where we note that the results differ because of a number of changes in the pipeline that were implemented. As already discussed in Sect.~\ref{sec:simulations} we changed the power law slope of the counts of faint galaxies, which shifts the bias as indicated by Fig.~9 in \citetalias{Hoekstra17}. We also improved the modelling of the PSF parameters: the pixel size of $0\farcs 1$ is relatively large compared to the FWHM of the PSF of a 1.2m diffraction limited telescope. In \citetalias{Hoekstra17} the correction for the PSF was based on parameters that were estimated directly from the poorly sampled images. Although this does not impact their main conclusions, it does change the actual biases. Here we use measurements of the PSF shear and smear polarisabilities \citep{KSB95,Hoekstra98} that were measured from $4\times$ oversampled images. Moreover, we increased the width of the weight function by a factor $1/\sqrt{\ln(2)}\approx 1.2$, which also changes the shear bias\footnote{We use the observed value of the half-light radius {\tt FLUX\_RADIUS} as measured by \sextractor\ to define the width of the weight function. For a Gaussian profile the corresponding dispersion $\sigma= {\tt FLUX\_RADIUS}/\sqrt{2\ln 2}$.}.

\begin{figure}
\centering
\leavevmode \hbox{%
  \includegraphics[width=8.5cm]{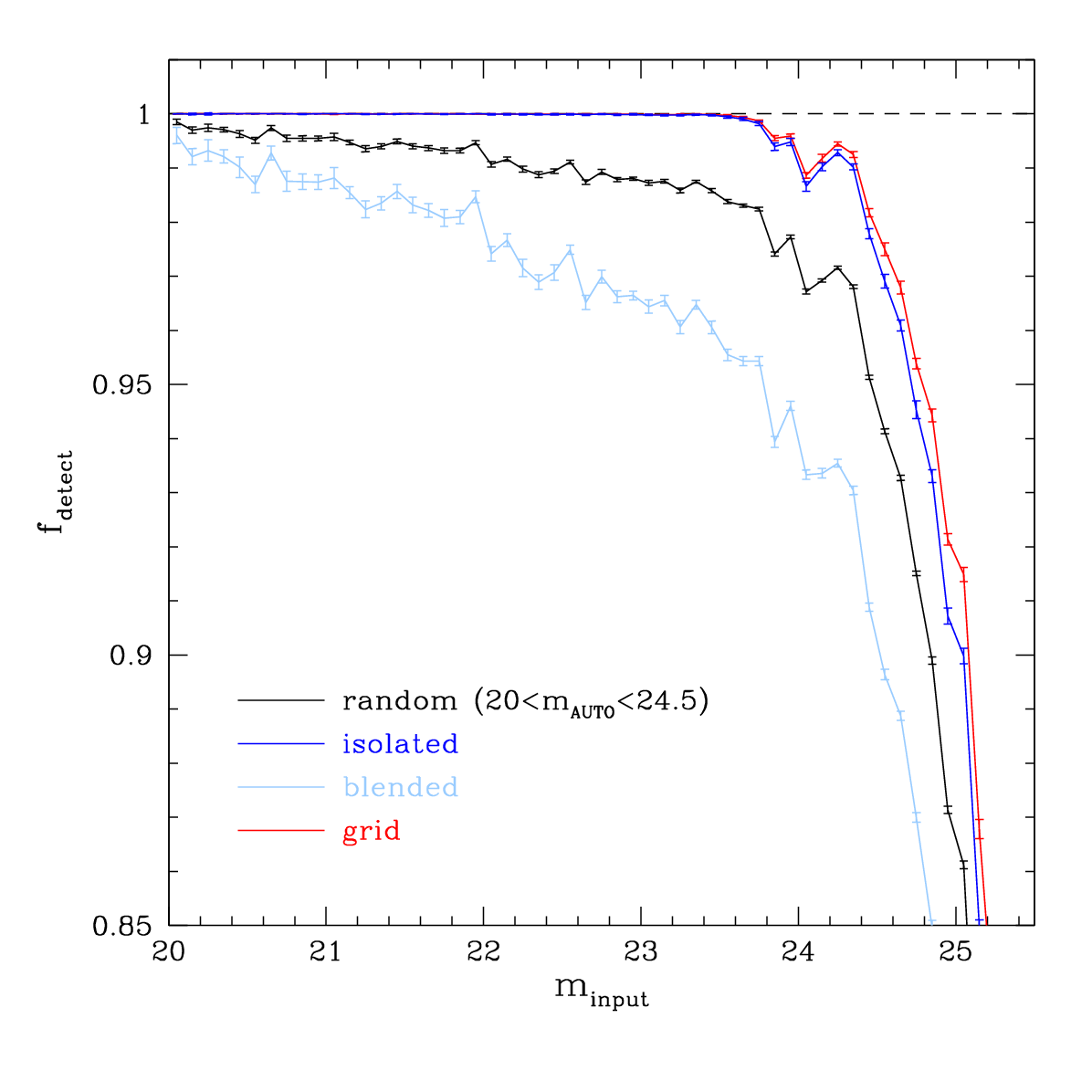}}
\caption{Fraction of the simulated galaxies that are detected by \sextractor\ as a function of the input magnitude, $m_{\rm input}$. The black line corresponds to the reference case where galaxies are placed randomly in the images. The blue line shows results for `isolated' galaxies, with a nearest neighbour more than $5''$ away, whereas the light blue line is for galaxies with a nearest neighbour within $2''$. In the latter case the fraction of detected galaxies is considerably lower, whereas the results for the `isolated' galaxies approaches that of the simulations where galaxies are placed on a grid about $9''$ apart (red lines). The error bars indicate the scatter in the results, and the lines connect the points.
  \label{fig:detfrac_magin}}
\end{figure}

\subsection{Detection and photometry performance}
\label{sec:performance}

Figure~\ref{fig:detfrac_magin} shows the fraction of simulated galaxies that were detected by \sextractor\ as a function of the  input magnitude, $m_{\rm input}$. To obtain this result we matched the input catalog to the \sextractor\ output and selected those objects that were detected within a radius of 3 pixels from the input coordinate. The black line shows the results for our baseline simulation, whereas the red line shows the fraction of detected objects if the galaxies are placed on a grid about $9''$ apart. In the latter case the sample of detected galaxies is complete down to $m_{\rm input}=23.5$, after which the completeness starts decreasing. The sample of galaxies detected in the baseline simulation is incomplete at all magnitudes, although 98\% of the galaxies are detected down to $m_{\rm input}=23.5$. The increased incompleteness is caused by blending, because the results for galaxies that have a nearest neighbour with $m_{\rm input}<26$ that is at least $5''$ away (blue line) resemble that of the grid-based images. If we instead select galaxies with a nearest neighbour with $m_{\rm input}<26$ within $2''$, the incompleteness increases (light blue line).

\begin{figure}
\centering
\leavevmode \hbox{%
  \includegraphics[width=8.5cm]{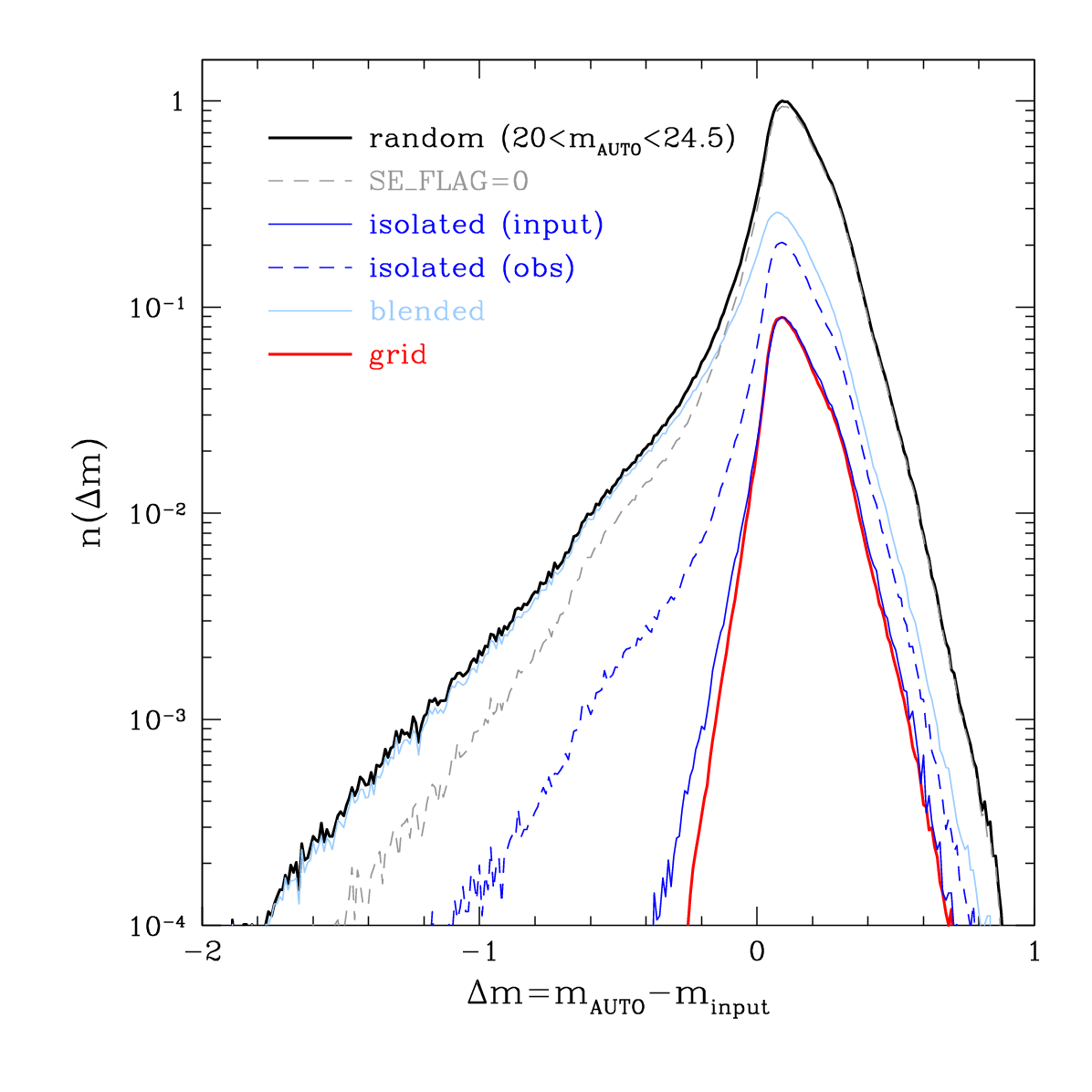}}
\caption{Distribution of $\Delta m$, the difference between $m_{\rm AUTO}$, the magnitude reported by \sextractor, and the input magnitude $m_{\rm input}$ for detected galaxies with $20<m_{\rm AUTO}<24.5$ for our baseline setup (solid black line; galaxies placed randomly). The distribution of `isolated' galaxies (solid blue line) matches that of the grid-based results (red line), whereas the tail toward negative $\Delta m$ matches that of `blended' galaxies. The light grey dashed line shows that many of the objects flagged by \sextractor\ are indeed blends, but that many remain undetected.
Blends even occur for objects that have no detected neighbour within $5''$ (dashed blue line).
  \label{fig:delta_mag}}
\end{figure}

This basic result shows that the detection of galaxies is affected by the presence of neighbouring galaxies. Before we proceed to explore the impact on shape measurements, we briefly examine the impact on the recovered magnitudes. 
The black line in Fig.~\ref{fig:delta_mag} shows the distribution of $\Delta m$, the difference between $m_{\rm AUTO}$, the magnitude reported by \sextractor\ as {\tt MAG\_AUTO}, and the input magnitude $m_{\rm input}$, for galaxies with $20<m_{\rm AUTO}<24.5$ in the baseline simulations. The results show a clear tail towards negative $\Delta m$, which is what we expect for blended objects. This is confirmed if we consider the distributions for `isolated' galaxies (blue; nearest neighbour $>5''$ away) and `blended' galaxies (light blue; nearest neighbour $<2''$ away): the distribution of isolated galaxies roughly matches that of the grid-based simulation (red line; normalisation matched to the blue curve), whereas the distribution of the blended galaxies, comprising 36\% of the galaxies, matches the tail for $\Delta m<-0.5$. 

The fraction of isolated galaxies is small, only 7.5\% of the galaxies match the criterion. In practice, however,  \sextractor\ will miss nearby neighbours if they are too close. If we use the distance to the nearest {\it detected} galaxy for the isolation criterion instead, we find that the fraction of apparently isolated galaxies is almost 19\%; the dashed blue line in Fig.~\ref{fig:delta_mag} shows the corresponding distribution, indicating the increased fraction of blends. 
Finally, \sextractor\ raises a flag for objects that if finds to be blended. The light grey dashed line in Fig.~\ref{fig:delta_mag} shows that it can indeed eliminate some of the blended objects, but many remain. Undetected blends are likely to bias the photometric redshifts, coupling these to biases in the shape measurements, but exploring this further is beyond the scope of this paper. 

Finally we note that the distributions do not peak around $\Delta m=0$, but that $\langle\Delta m\rangle=0.14$ for $20<m_{\rm AUTO}<24.5$ in the grid-based simulations. The amount of missing flux does depend somewhat on the brightness, increasing from $\langle \Delta m\rangle=0.056$ for the brightest galaxies ($m_{\rm AUTO}=20$) to $\langle \Delta m\rangle=0.166$ for the faintest ones ($m_{\rm AUTO}=24.5$). It also depends somewhat on the source ellipticity, which partly explains the asymmetry towards positive values of $\Delta m$.  Although the dependence of 
$\Delta m$ on ellipticity is modest, it implies that a simple magnitude cut may lead to changes in the ellipticity distributions of the detected galaxies, potentially complicating the link between shape measurements and photometric redshift determinations further.

\section{Detection bias}
\label{sec:detection}

The measurement of the weak gravitational lensing signal relies on accurate estimates of the shapes of distant galaxies, which are both faint and small. The images are corrupted by noise and instrumental effects. It is essential to remove, or at least account for, these sources of bias. For this reason most effort has focused on undoing the biases in the shape measurement step itself, but the preceding step, the detection (and selection) of galaxies that are used in the analysis, has received much less attention. 

\henk{As shown already in \cite{Hirata03} we do expect the detection of objects to introduce a bias. Gravitational lensing conserves the surface brightness, and as a result a galaxy with an intrinsic orientation perpendicular to the shear will appear rounder at the same surface brightness level. Since \sextractor\ uses a surface brightness threshold and a circular kernel for the detection, such a galaxy is more likely to be detected, resulting in the average shear to be biased low.} The detection and selection biases are typically much smaller than the shape measurement biases, but they can no longer be ignored for Stage IV surveys \citep{detf}, and require more detailed study \citep[as shown by][they are already relevant for stage III surveys]{FC17,Kannawadi19}.

We will discuss both detection and selection biases. The former refers to the very first step in the analysis, resulting in a sample of objects for which a shape measurement can be attempted. The subsequent shape measurement may not always be successful, or different weights may be assigned to the measurement, which leads to selection biases. Similarly the desire to divide the galaxies into tomographic bins introduces selection biases that need to be accounted for. We emphasise that these biases occur even if the shape measurement itself is unbiased.

To mimic a perfect shape measurement we follow \cite{FC17} and compute the true measured ellipticity based on the input complex ellipticity $\epsilon^{\rm s}$ and applied complex shear $\gamma$ as given by Eq.~(\ref{eq:ellipticity}). For each galaxy detected by \sextractor\ we find the nearest input galaxy. For the analysis we consider only galaxies with observed magnitudes $m_{\rm AUTO}<25$, but the input catalogue includes many more galaxies that are fainter. As most of those are not detectable individually (see e.g. Fig.~\ref{fig:detfrac_magin}), we only consider the nearest object with $m_{\rm input}<26$ from the input catalogue. We define a mismatch if the separation is more than 3 pixels, which is the case for 0.2\% of the objects with $m_{\rm AUTO}<25$. The fraction is larger for fainter objects (e.g., 1.4\% for detections with $25<m_{\rm AUTO}<26$) suggesting that some of these are just noise peaks. However, we note that such misidentifications do not bias \henk{our} shear estimate, but rather introduce noise in our measurement because the shape noise is not cancelled in this case\footnote{\henk{In our case, a noise peak is still associated with an input galaxy, resulting in imperfect shape noise cancellation only. In contrast, including noise peaks in an actual cosmic shear analysis does lower the signal. In practice, however, requiring robust photometric redshifts using multi-band observations will remove most, if not all, of these.}}. Even though the impact of these mismatches on the results is negligible, we omit them from our analysis.

\begin{figure}
\centering
\leavevmode \hbox{%
  \includegraphics[width=8.5cm]{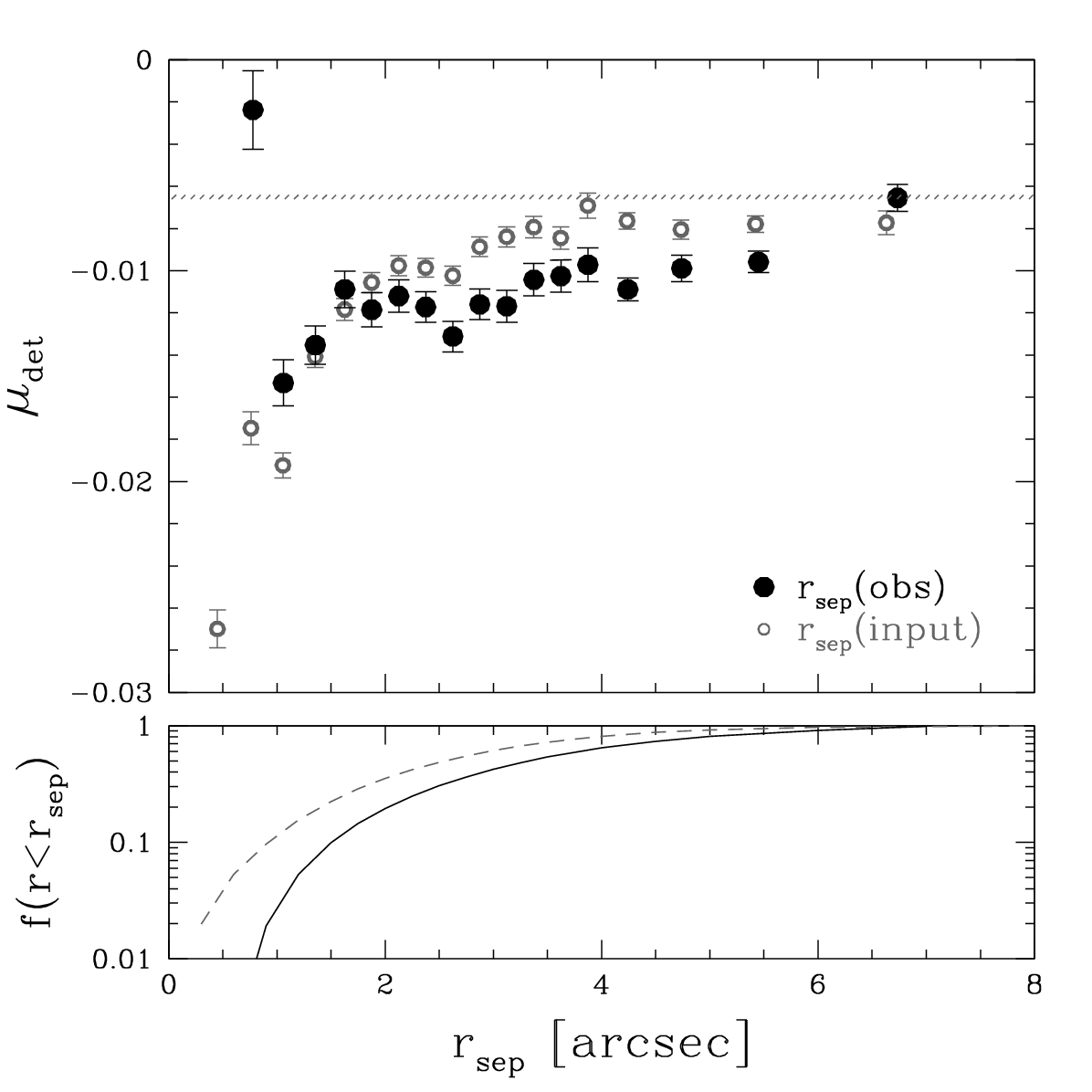}}
\caption{{\it Top panel:} Detection bias for galaxies with $20<m_{\rm AUTO}<24.5$ as a function of $r_{\rm sep}$, the distance to the nearest object detected by \sextractor\ (black points). The open grey points show the detection bias as a function of the nearest neighbour in the input catalogue brighter than $m_{\rm input}=26$ (grey open points). \henk{For reference, the hatched region indicates the detection bias for the grid-based simulations}; {\it Bottom panel:} The fraction of galaxies that have a neighbour within a distance $<r_{\rm sep}$ in the input catalogue (grey dashed line) or detection catalogue (black line). For small separations many of the true blends are not recognised as such.
  \label{fig:detbias_separation}}
\end{figure}

\begin{table*}
  \caption{Average multiplicative and additive biases for galaxies with $20<m_{\rm AUTO}<24.5$.\label{tab:bias}}
  \centering
\begin{tabular}{lrrrr}
  \hline
  \hline
   setup &  $\mu_1$ & $\mu_2$  & $c_1~[\times 10^5]$ & $c_2~[\times 10^5]$ \\
   \hline
   \multicolumn{5}{c}{\bf baseline}\\
   \hline
   \sextractor\,  & $-0.010\,61\pm 0.000\,13$ & $-0.010\,53\pm0.000\,13$ & $-0.73\pm0.51$ & $-0.73\pm0.49$ \\	
   \sextractor\, -- no background & $-0.010\,43\pm0.000\,18$ & $-0.009\,74\pm0.000\,18$ & $-0.68\pm 0.73$ & $0.36\pm0.71$\\
   \sextractor\, -- {\tt FLAG}=0 & $-0.012\,14\pm0.000\,23$ & $-0.011\,78\pm0.000\,22$ & $-0.71\pm 0.93$ & $0.70\pm0.92$\\      
   KSB detection & $-0.019\,17\pm0.000\,23$ & $-0.018\,72\pm 0.000\,23$ & $-0.09\pm0.88$ & $1.17\pm0.88$\\
   KSB selection & $-0.018\,66\pm0.000\,22$ & $-0.018\,19\pm0.000\,22$ & $-0.08\pm 0.86$ & $1.12\pm0.86$ \\
   KSB shapes & $-0.089\,15\pm0.000\,31$ & $-0.087\,57\pm0.000\,32$ & $-3.34\pm 1.20$ & $0.74\pm 1.21$ \\   
   \hline
      \multicolumn{5}{c}{\bf sheared image}\\
   \hline
   \sextractor\, & $-0.010\,62\pm0.000\,19$ & $-0.010\,43\pm0.000\,18$ & $-0.84\pm0.71$ & $-0.18\pm0.73$\\
   KSB detection & $-0.019\,28\pm0.000\,22$ & $-0.019\,01\pm 0.000\,23$ & $-0.19\pm0.88$ & $-1.18\pm0.87$\\
   KSB selection & $-0.018\,77\pm0.000\,22$ & $-0.018\,47\pm 0.000\,22$ & $-0.16\pm 0.86$ & $-1.18\pm0.85$\\
   KSB shapes & $-0.088\,79\pm0.000\,31$ & $-0.087\,66\pm0.000\,33$ & $-2.14\pm1.21$ & $-1.34\pm1.22$\\
     \hline
   \multicolumn{5}{c}{\bf grid} \\
   \hline
   \sextractor\,	 	& $-0.006\,94\pm0.000\,24$ & $-0.006\,97\pm0.000\,24$ & $-1.53\pm0.9$   & $0.85\pm0.91$ \\
   KSB detection 	& $-0.013\,25\pm0.000\,28$ & $-0.012\,57\pm0.000\,29$ & $-0.81\pm1.15$ & $-2.83\pm1.12$\\
   KSB selection	& $-0.012\,81\pm0.000\,27$ & $-0.012\,09\pm0.000\,29$ & $-0.79\pm1.11$ & $-2.77\pm 1.08$ \\
   KSB shapes		& $-0.053\,57\pm0.000\,30$ & $-0.052\,51\pm0.000\,32$ & $-1.14\pm1.25$ & $-1.80\pm1.25$ \\
  \hline
  \hline  
\end{tabular}
\bigskip
\begin{minipage}{\linewidth}
 \tablefoot{In the `baseline' case the galaxies are placed randomly and their images are sheared. For the `sheared image' results the full scene is sheared instead, thus altering the positions. Galaxies are placed on a regular grid, about $9''$ apart for the `grid' results. The rows labelled `\sextractor' report the detection bias. The biases for objects with a KSB shape measurement are labelled `KSB detection', and as `KSB selection' when the weighting scheme is included. The results using the actual KSB shape measurements are reported as `KSB shapes'. \henk{All the KSB measurements also include the \sextractor\ detection bias.} The reported uncertainties may differ for similar setups, because fewer simulations were analysed.}
\end{minipage}
\end{table*}

More important are the cases where the object is blended with a neighbouring one, which can also lead to a shift in the location of the detection. In $0.4\%$ of the detections with $m_{\rm AUTO}<25$ we identify a brighter object in the input catalogue within a radius of 3 pixels. As the galaxies are placed randomly, these are mere chance projections, which is consistent with the observed distribution of separations. In these cases we assign the input properties of the brighter object, because a shape measurement algorithm would be more sensitive to its surface brightness distribution. 

We then proceed to compute the shear biases by comparing the average ellipticity of the detected galaxies to the input shear $\gamma^{\rm true}_i$ (where the index $i \in \{1,2\}$ corresponds to the real or imaginary part of the shear, respectively). The former is an estimate of the shear, as can be seen by averaging Eq.~(\ref{eq:ellipticity}): $\langle \epsilon^{\rm obs}_i\rangle=\gamma_i^{\rm obs}$. As is common, we assume that the observed shear and true shear are related as:
\begin{equation}
\gamma_i^{\rm obs}=(1+\mu_i)\gamma_i^{\rm true}+c_i,
\end{equation}
where $\mu_i$ is the multiplicative shear bias, and $c_i$ is the additive shear bias. The values for $\mu_i$ are expected to be very similar \citep{Kitching19}. We determine both components separately, and if they are  consistent we refer to $\mu$ as the average of the two components. Finally, we note that because we create pairs of images where the galaxies are rotated by $90^{\circ}$ the presence of a bias means that one of the two images is not detected, or assigned a magnitude such that it is not included, and that the probability of detection depends on the applied shear itself.

\subsection{Detection bias estimates}
\label{sec:estimates}

Figure~\ref{fig:detfrac_magin} shows that the presence of neighbouring galaxies affects the ability of {\tt SExtractor} to detect galaxies. We now proceed to explore whether this results in a bias in the shear. Unless specified otherwise we report biases for galaxies with $20<m_{\rm AUTO}<24.5$, which was adopted by \citetalias{Hoekstra17} as a good approximation for the range used by \Euclid. This allows for a direct comparison to their results for the overall shear bias, although we note that our analysis differs somewhat (see Sect.~\ref{sec:setup} for details). We present results for different setups in Table~\ref{tab:bias}.

For our baseline setup, where galaxies are placed randomly, we measure $\mu^{\rm det}_1=-0.010\,61\pm 0.000\,13$ and $\mu^{\rm det}_2=-0.010\,53\pm0.000\,13$, where the uncertainties reflect the finite number of images that were analysed. We do not detect a significant additive bias, but the detection bias is significant for our \Euclid-like setup, especially if we contrast this with the overall requirement that $|\mu|<2\times 10^{-3}$ \citep{Cropper13}. Both multiplicative shear biases agree 
($\langle\mu^{\rm det}_1-\mu^{\rm det}_2\rangle=(-0.9\pm1.8)\times 10^{-4}$), which is why we show the average of both components in most figures. In Table~\ref{tab:bias} we also present the detection bias when we fix the background to its true value (i.e. zero; reported as `no background'). The changes in multiplicative shear bias are small, but significant\footnote{\henk{The measurements for different scenarios are based on the same images, and are therefore correlated. We account for this by computing the difference first and reporting its statistics. As a result, the difference may be determined more precisely than the bias itself.}}: $\Delta\mu_1=0.000\,37\pm0.000\,11$ and $\Delta\mu_2=0.000\,62\pm0.000\,11$   

In the baseline setup we do not shear the full scene, but only shear the galaxy images. In reality the shearing also alters the positions, which in turn might affect the results as the separations between neighbouring objects change slightly. 
If we shear the full image instead, the difference with respect to the baseline case where we only shear the galaxy images is  $\Delta\mu=(0.52\pm1.59)\times 10^{-4}$, i.e. we do not observe a significant difference. Similarly the additive biases are consistent with the baseline results. To obtain this estimate we used the fact that the galaxy images are the same for both setups \henk{(though not their positions)}, but that the background noise realisation is slightly different.

\henk{The black points in Fig.~\ref{fig:detbias_separation} show the detection bias for galaxies with $20<m_{\rm AUTO}<24.5$ as a function of $r_{\rm sep}$, the distance to the nearest object {\it detected} by \sextractor. For large separation, the bias approaches the average bias we measured for the grid-based simulations (indicated by the hatched region), but is typically larger. This is because not all blends are identified as such. For reference we also show the bias as a function of the distance to the nearest neighbour in the input catalogue brighter than $m_{\rm input}=26$ (grey open points). The amplitude of the bias changes rapidly for galaxies with $r_{\rm sep}<1''$, and such galaxies are probably best omitted from the analysis. The bottom panel in Fig.~\ref{fig:detbias_separation} shows that this applies to about 10\% of the galaxies. In reality this number will be higher because of clustering \citep{Martinet19}.}

\begin{figure*}
\centering
\leavevmode \hbox{%
  \includegraphics[width=8.5cm]{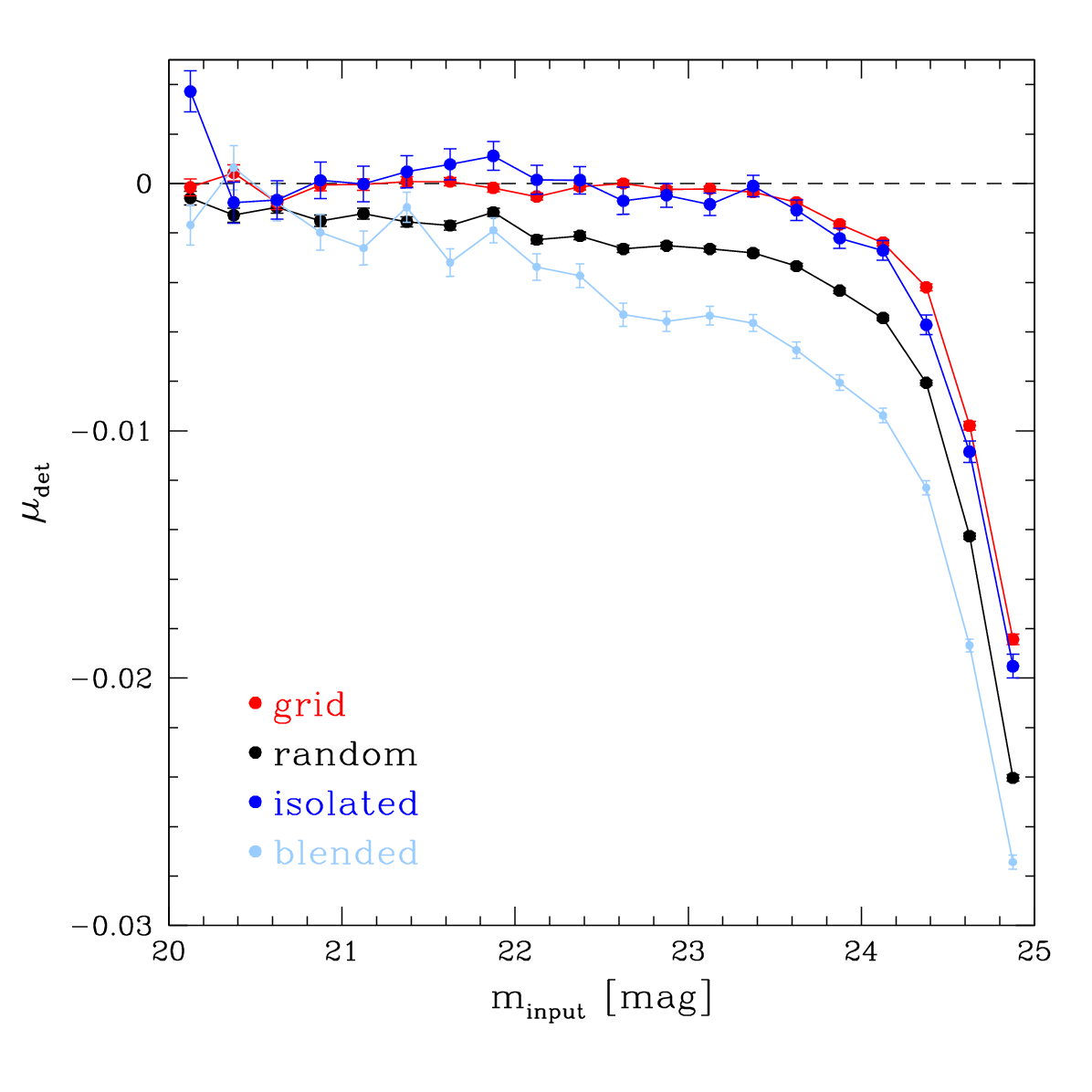}
    \includegraphics[width=8.5cm]{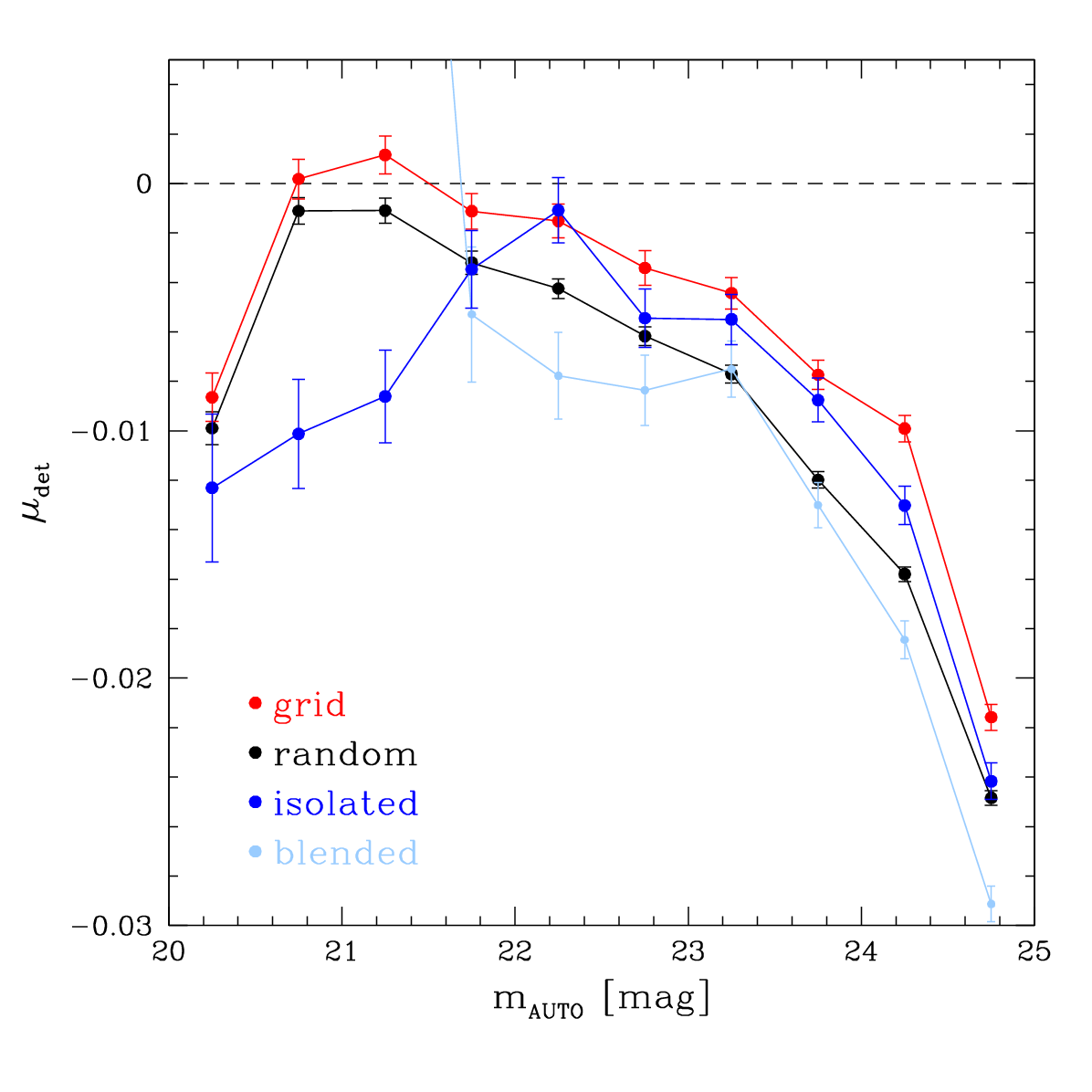}}
\caption{{\it Left panel:} Multiplicative detection bias $\mu_{\rm det}$ as a function of the input apparent magnitude when galaxies are placed on a grid (red points) or placed randomly (black points). The blue points show the results for isolated galaxies where the nearest neighbour is more than $5''$ away, whereas the light blue points show the detection bias for galaxies with a neighbour within $2''$ (blended). {\it Right panel:} Multiplicative detection bias as a function of observed properties. The classification into isolated and blended galaxies is based on the nearest \emph{detected} galaxy in this case.
The lines connect the points to show the behaviour for the different samples more clearly. The bias for the bright blended galaxies is beyond the axis limits of the chart. 
  \label{fig:detbias_mag}}
\end{figure*}

As indicated by Fig.~\ref{fig:delta_mag} selecting objects with  \sextractor\ {\tt FLAG}$=0$ reduces the occurrence of blends, and we expect the detection bias to be reduced (see Fig.~\ref{fig:detbias_separation}). Instead we find that the bias is increased by about 13\%, implying that the flagging of blended objects is actually done in a shear dependent fashion.

These results indicate that blending is a significant source of detection bias that depends significantly on the local galaxy density. We note, however, that the bias does not vanish for large separations, but rather converges to the bias we obtained for our grid-based simulations, indicated by the hatched horizontal region (and reported in Table~\ref{tab:bias}), if we select galaxies based on the distance to the nearest neighbour in the input catalogue. In the more realistic case (open grey points), where we separate galaxies based on the distance to the nearest detected galaxy, the bias is even larger because many blends remain undetected. 

To investigate this further we show $\mu_{\rm det}$ as a function of magnitude
in Fig.~\ref{fig:detbias_mag}. The left panel, where we show results as a function of the input magnitude, $m_{\rm input}$, is the shear detection bias equivalent of Fig.~\ref{fig:detfrac_magin}. In this case the shape noise cancellation results in small uncertainties, because galaxies are included in the correct magnitude bin by design. The shear bias arises because the probability of detecting faint galaxies is affected by the orientation of the galaxy with respect to the applied shear: galaxies that are aligned perpendicular to the shear are more likely to be detected. The bias is negligible for bright galaxies, and thus can be reduced by increasing the depth of the observations, something we will explore further in Sect.~\ref{sec:noise}.

Similar to Fig.~\ref{fig:detfrac_magin} we find that the bias for isolated galaxies ($r^{\rm in}_{\rm sep}> 5''$) matches that of the grid-based images, whereas the bias is larger for blended galaxies ($r^{\rm in}_{\rm sep}<2''$). Comparison of the biases reported in Table~\ref{tab:bias} suggests that both blending and the shear-dependent detection probability are important. The bias at bright magnitudes is caused by blending, whereas for fainter galaxies the detection probability itself depends on the orientation with respect to the applied shear. 

In reality the situation is complicated by the fact that the observed magnitudes are affected by blending, the applied shear, and measurement uncertainties, all of which spread the biases over a wider range in magnitudes and lead to larger uncertainties owing to imperfect shape noise cancellation. Consequently, the error bars in the right panel of Fig.~\ref{fig:detbias_mag} are increased, and the detection bias affects a larger range in magnitude. \henk{In particular, as shown by the asymmetric distribution of magnitude errors in Fig.~\ref{fig:delta_mag}, blending scatters objects towards a brighter magnitude bin. Such blends are not always identified, and can thus introduce significant biases even for apparently bright galaxies. For instance, the bias for the bright blended galaxies is far beyond the axis limits of the chart.  We also caution that the results for the brightest magnitude bin suffer from extreme Eddington bias, because our input catalogue does not include galaxies brighter than $m=20$.}

\begin{figure}
\centering
\leavevmode \hbox{%
  \includegraphics[width=8.5cm]{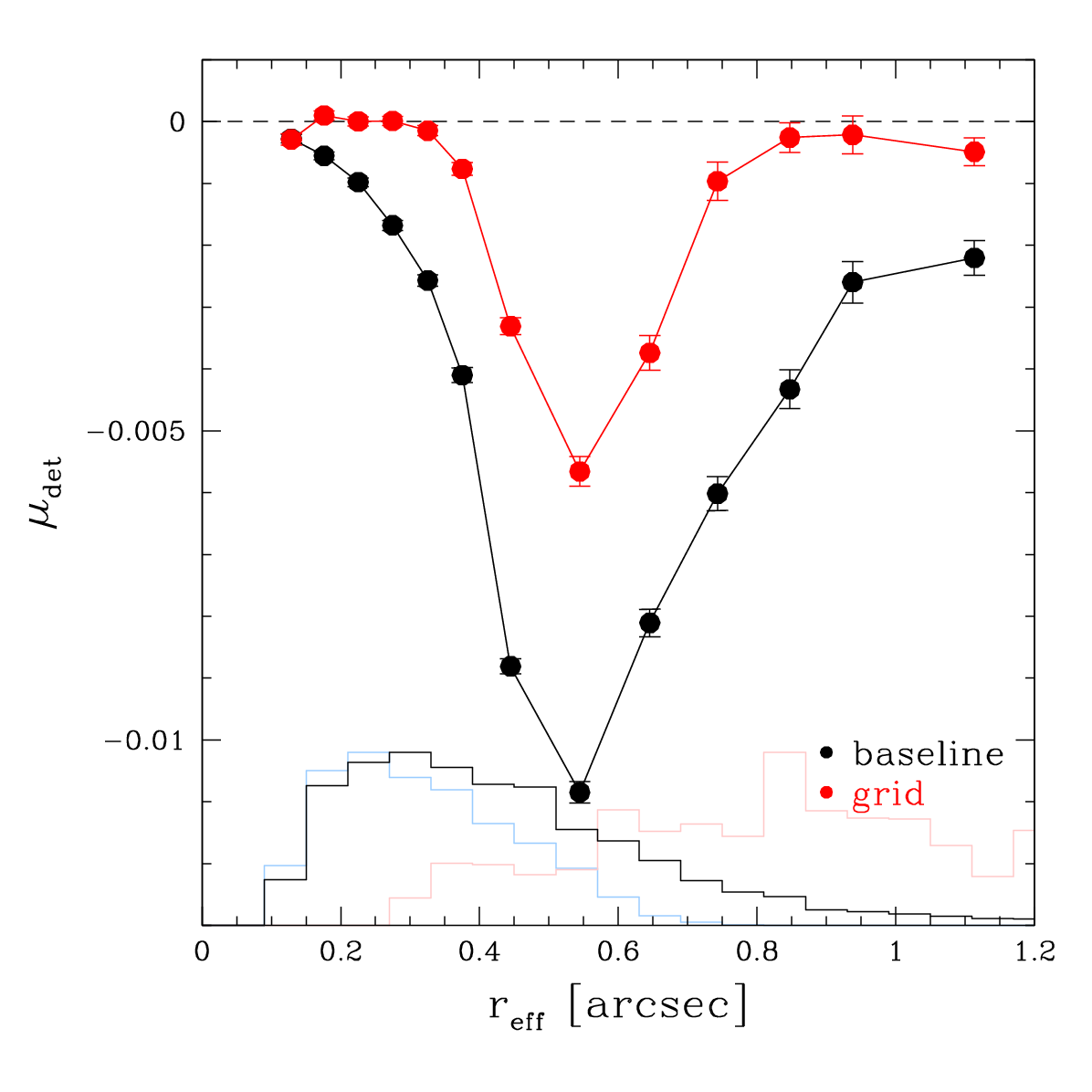}}
\caption{Multiplicative detection bias $\mu_{\rm det}$ as a function of the
input half-light radius, $r_{\rm eff}$, for  galaxies with $20<m_{\rm AUTO}<24.5$. The black and red lines correspond to the baseline and grid-based cases, respectively. The histograms show the distributions of galaxy sizes
(black: all galaxies; red: $m_{\rm AUTO}<21$; blue: $24<m_{\rm AUTO}<24.5$)
The observed behaviour is the result of the change in size as a function of brightness.
\label{fig:bias_size}}
\end{figure}

Figure~\ref{fig:bias_size} shows the multiplicative detection bias as a function of the input half-light radius ($r_{\rm eff}$) for the baseline (black) and grid-based (red) simulations. For both cases we observe a strong dependence with galaxy size, which is the combined result from the underlying distribution of fluxes and the correlation between size and brightness. After all brighter galaxies are more likely to be detected, whilst for a given flux a smaller galaxy is detected with a higher significance. The latter drives the increase in detection bias with increasing $r_{\rm eff}$, but as the mean brightness increases with increasing size, the probability of detection increases once more. 
\henk{Comparison of the bias as a function of $r_{\rm eff}$ for isolated galaxies with the grid-based results show that they agree well.} Hence,  the difference between the grid-based and baseline simulations is caused by blending, which affects galaxies of all sizes. 

In contrast to what was done in Fig.~\ref{fig:detbias_mag}, we do not show the bias as a function of {\tt FLUX\_RADIUS}, the  half-light radius determined by \sextractor, because it correlates with ellipticity. Consequently, a split by {\tt FLUX\_RADIUS} is an implicit selection in ellipticity, resulting in large biases. If one wants to split the source sample by a particular observable, it is important to verify that it does not correlate with input ellipticity. \henk{This may not be fully feasible in practice, but at least one should aim to minimise the dependence. Interestingly, we find that {\tt MAG\_AUTO} only weakly correlates with the input ellipticity. This suggests that splitting the sample into tomographic bins based on magnitude and colour may not increase the selection bias much, although further study would be required to quantify this.}

\subsection{Dependence on noise level}
\label{sec:noise}

\begin{figure}
\centering
\leavevmode \hbox{%
  \includegraphics[width=8.5cm]{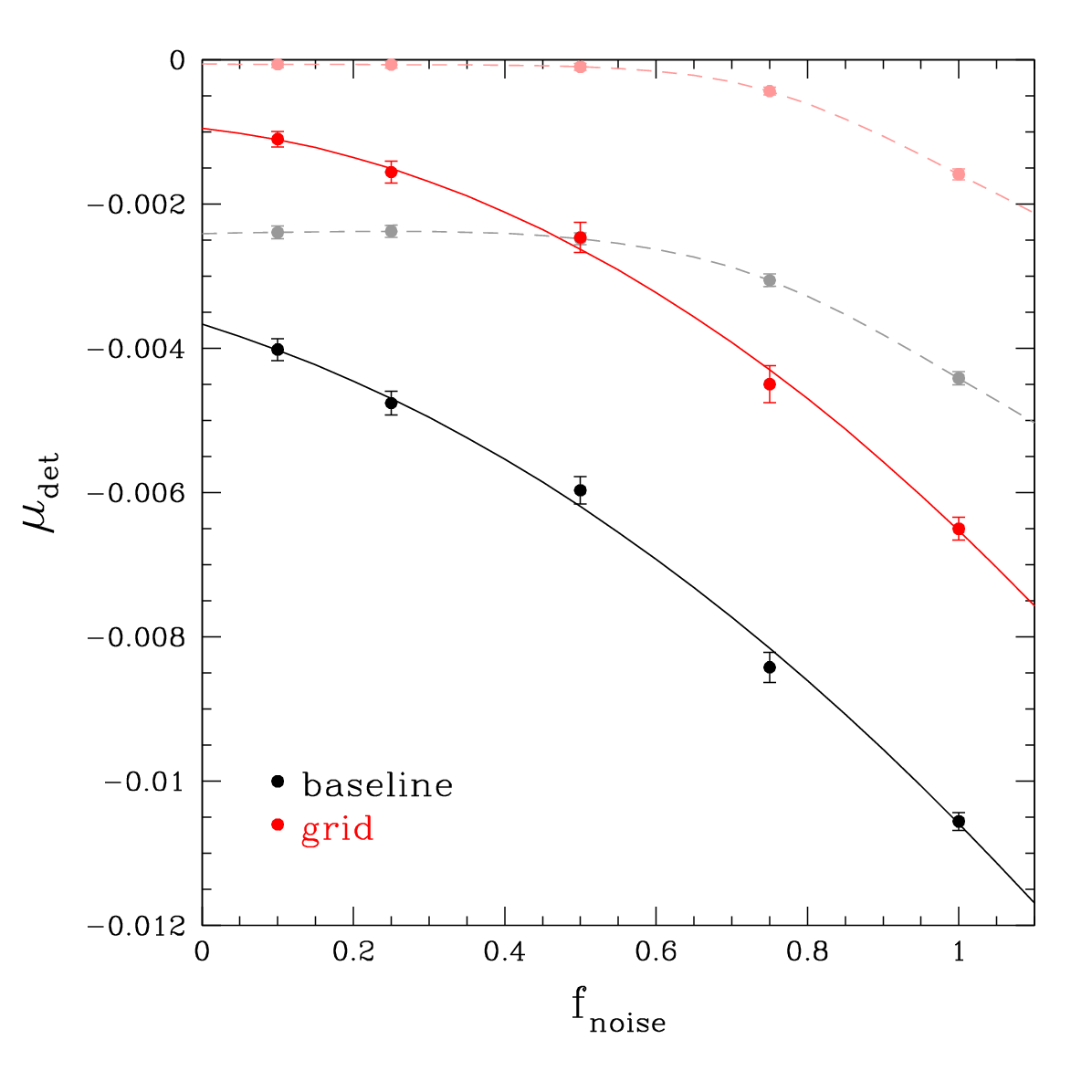}}
\caption{Multiplicative detection bias $\mu_{\rm det}$ as a function of the background noise level, which is multiplied by a factor $f_{\rm noise}$ with respect to the baseline case. The black and red lines correspond to the baseline and grid-based cases, respectively. The solid lines show results for  galaxies with $20<m_{\rm AUTO}<24.5$, whereas the (light-coloured) dashed lines indicate the bias if we select using the input magnitudes, $20<m_{\rm input}<24.5$. In the latter case the bias vanishes for the grid-based case as the noise level is low, but for the baseline case the bias plateaus to $\mu_{\rm det}=-0.0024$ as a result of blending.\label{fig:bias_noise}}
\end{figure}

Figure~\ref{fig:detbias_mag} shows that the detection bias is negligible for bright, isolated galaxies. Hence, we expect that the detection bias can be reduced by obtaining deeper data. The results in Fig.~\ref{fig:bias_noise} show that this is indeed the case: it shows the multiplicative detection bias when the noise level in the image is multiplied by $f_{\rm noise}$ (where $f_{\rm noise}=1$ corresponds to the baseline case). The black (red) points  show the results for the baseline (grid) simulations for galaxies with $20<m_{\rm AUTO}<24.5$. These are well fit by a second order polynomial (solid lines). 

The average increase in detection bias of $\mu_{\rm base}-\mu_{\rm grid}=-0.0035$ is caused by blending and increases only weakly with increasing noise level. Moreover, even for low noise levels blending leads to a floor in the detection bias that is about $\sim -0.004$. Interestingly, the bias does not completely vanish in the grid-based simulations at low noise levels. This is the result of our galaxy selection which is based on the magnitude estimates by
\sextractor. If we instead select the galaxies based on their true (but unobservable) magnitudes the bias quickly vanishes (light red points and red dashed line). This implies that the estimate of $m_{\rm AUTO}$ depends slightly on the shear. For the baseline case (light grey points) the bias plateaus to  $\mu_{\rm det}=-0.0024$ as a result of blending.

\henk{These results show that the detection bias is a combination of blending and the sample selection (in our case a  magnitude cut).
Although we find that it may be possible to reduce detection bias somewhat using deeper observations, blending quickly becomes a limiting factor, even in space-based data.}

\subsection{KSB biases}
\label{sec:KSB}

In Table~\ref{tab:bias} we also present measurements for the shear biases for the KSB algorithm \citep{KSB95, Hoekstra98}, because we made a number of changes in both the simulations and the measurement setup since \citetalias{Hoekstra17} (see Sect.~\ref{sec:simulations}). With this modified setup we measure
at total shear bias of $\mu_1^{\rm KSB}=-0.089\,15\pm0.000\,31$ and $\mu_2^{\rm KSB}-0.087\,57\pm0.000\,32$. The results also suggest that a small additive bias is introduced, although more simulations would be needed to confirm the result. The detection bias is about 9 times smaller than the total shear bias, which explains the focus of previous studies on shear bias. 

We also report the biases introduced by the steps in the shape measurement analysis following the initial \sextractor\ detection. The ability of the KSB algorithm to measure a shape also depends on the shear, resulting in an increase in the detection bias. For the `KSB detection' bias we use the true shapes, but only for those galaxies where a shape was measured. The results in Table~\ref{tab:bias} show that the bias doubles for all image setups. The bias is reduced somewhat if we weight the true ellipticities with the KSB weights (`KSB selection').

Although the shape measurement bias itself is dominant, the detection bias is not negligible. As the detection bias is most readily quantified using image simulations, like the one we use here, we need to quantify the sensitivity of the detection bias to the simulation setup, similar to what was done by \citetalias{Hoekstra17} for the overall shear bias. We return to this in Sect.~\ref{sec:realism}, but first examine the sensitivity to the \sextractor\ setup and PSF anisotropy.

\subsection{Sensitivity to detection setup}
\label{sec:var_setup}

Table~\ref{tab:setup} lists the main parameters that play a role in the object detection. These can be grouped into three categories. The first three pertain to the detection itself, the next three affect the behaviour for blended objects, and the last two are relevant for the background estimation. As already mentioned, the background parameters do not play an important role for our study. Also the choices for {\tt DETECT\_MINAREA} and {\tt DETECT\_THRESH} do not affect our findings for galaxies with $20<m_{\rm AUTO}<24.5$ (provided they are not  modified significantly), but the choice of the filter that is used to detect objects is relevant. To detect objects in the presence of noise, the images are convolved with a suitable kernel before searching for peaks. The optimal filter has a profile that matches the object of interest. For this reason \cite{KSB95} developed a hierarchical peak finder, which employs a series of filters, but is slower. {\tt SExtractor} is run with a single filter, specified by the keyword {\tt FILTER\_NAME}. Here we run it using the various predefined round Gaussian filters, defined by their dispersion $\sigma_{\rm filter}$.

\begin{figure}
\centering
\leavevmode \hbox{%
  \includegraphics[width=8.5cm]{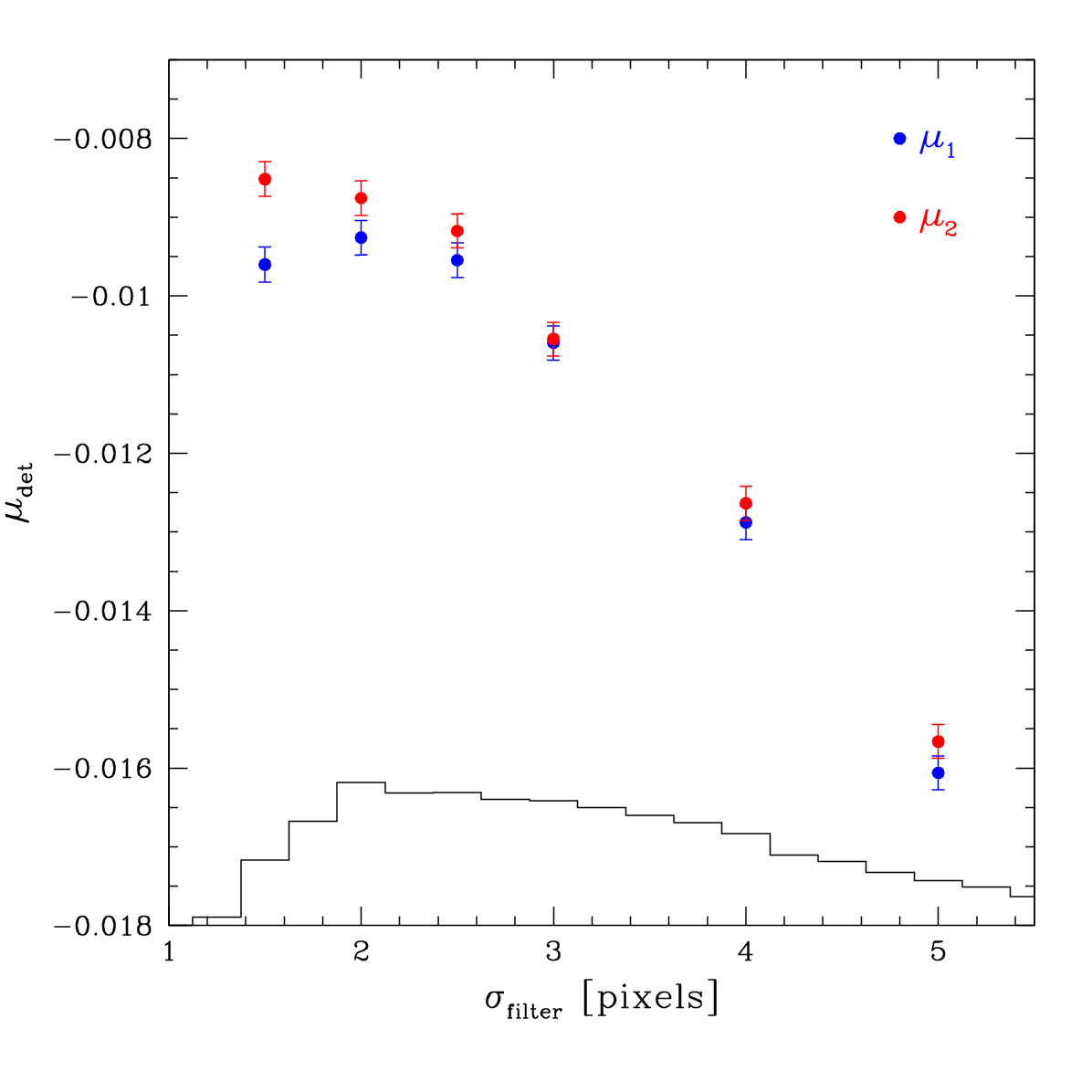}}
\caption{Multiplicative detection bias $\mu_{\rm det}$ as a function of the width of the filter used  in the detection step for galaxies with $20<m_{\rm AUTO}<24.5$. The blue (red) points correspond to $\mu_1$ ($\mu_2$). The histogram shows the distribution of corresponding sizes based on the half-light radius of the galaxies, suggesting that a width of $2-3$ pixels is best. The bias increases quickly for larger values of $\sigma_{\rm filter}$.
  \label{fig:detbias_filter}}
\end{figure}

The results are presented in Fig.~\ref{fig:detbias_filter} for galaxies with  $20<m_{\rm AUTO}<24.5$, where we show the multiplicative detection biases for the two shear components separately. They show a similar behaviour with filter width $\sigma_{\rm filter}$, but we observe a small offset, which is more significant for smaller filter sizes. The histogram shows the distribution of corresponding sizes based on the half-light radii of the galaxies, suggesting that using a Gaussian filter with a width of $2-3$ pixels is best. The bias increases quickly for larger values of $\sigma_{\rm filter}$.

Figures~\ref{fig:detfrac_magin} and~\ref{fig:detbias_mag} show that the presence of neighbouring objects affects the detection and introduces detection bias. We explore how changes in the parameters that  affect the deblending of objects ({\tt DEBLEND\_MINCONT} , {\tt DEBLEND\_NTHRESH}, and {\tt CLEAN\_PARAM}) in Appendix~\ref{app:setup}. We find that the detection biases for the default parameters are close to optimal, and that even substantial variations have only a minimal impact. Hence, the observed detection biases are not the result of a poorly chosen setup of \sextractor. 

\subsection{Sensitivity to PSF anisotropy}
\label{sec:psfan}

\begin{figure}
\centering
\leavevmode \hbox{%
  \includegraphics[width=8.5cm]{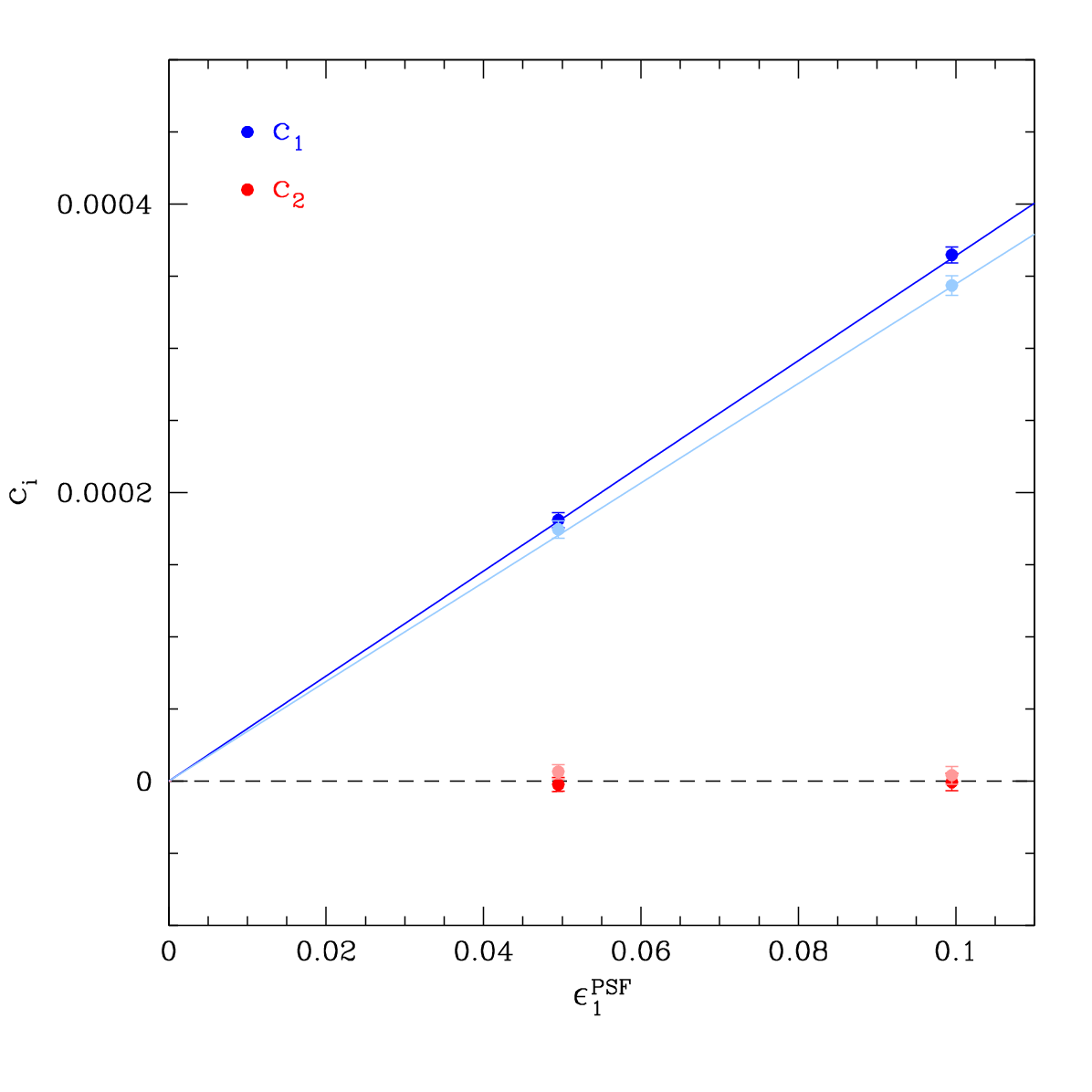}}
\caption{Additive bias $c_1$ (blue) and $c_2$ (red) as a function of the PSF ellipticity $\epsilon_1^{\rm PSF}$. 
The bright colours correspond to the baseline case where galaxies are placed randomly, whereas the light coloured points were obtained by placing galaxies on a grid. In the former case the additive detection bias is about 5.6\% higher,
but in both causes galaxies are preferentially detected when their orientation is aligned with the PSF. We do not observe a significant $c_2$ (red points), nor a change in multiplicative bias (not shown).
  \label{fig:cbias_epsf}}
\end{figure}

Thus far we focused only on the multiplicative detection bias that arises because the probability of detecting a galaxy depends on its orientation with respect to the shear \citep{Hirata03}. However, we expect the PSF to be anisotropic due to optical aberrations that are practically unavoidable, especially for a wide field imager. Such PSF anisotropy also introduces a preferred direction. \henk{In this case surface brightness is not conserved, and a galaxy with an intrinsic orientation parallel to the PSF ellipticity direction will have a higher peak brightness compared to a galaxy oriented orthogonal to the PSF anisotropy. As a consequence, we expect to preferentially detect galaxies that are aligned with the PSF anisotropy, leading to a positive additive bias \citep{Kaiser00,Bernstein02}.}

To study this, we created simulated images where the PSF was made elliptical in the $\epsilon_1$ direction and ran \sextractor\ to quantify the additive and multiplicative shear biases. Figure~\ref{fig:cbias_epsf} shows the resulting additive bias $c_i$. We find that that $c_2$ is consistent with zero (red and light red points), but we find that the object detection introduces a significant additive shear bias $c_1$, both when galaxies are placed on a grid (light blue points) or placed randomly (blue points); the bias in the latter case is only 5.6\% higher. 

\henk{As expected,} the bias has the same sign as the PSF anisotropy, demonstrating that \sextractor\ preferentially selects objects that are aligned with the PSF \citep[this was also observed in][]{Kannawadi19}. Although the amplitude is small, only $0.4\%$ of the original PSF ellipticity, this bias cannot be ignored if the PSF is anisotropic. For instance, \cite{Cropper13} argue that $|c|<5\times 10^{-4}$ is required, which is reached for $\epsilon^{\rm PSF}=0.137$. \henk{PSF anisotropy is therefore a non-negligible source of additive detection bias, which will vary spatially because we expect the PSF ellipticity to change across the field-of-view.}

We also examined the change in multiplicative shear bias as a function of $\epsilon^{\rm PSF}$ and we found no significant trend. This is worth noting, because we show in Appendix~\ref{app:additive} that sources of additive bias tend to introduce multiplicative biases of similar amplitude, but opposite sign in shape measurements. This connection can be used to empirically estimate the level of multiplicative bias for (residual) systematic effects that cause additive bias. In contrast, the lack of a change in multiplicative detection bias in the case of an anisotropic PSF shows that detection bias is fundamentally different from the shape measurement process itself. 

\section{Realism of the simulations}
\label{sec:realism}

\begin{figure}
\centering
\leavevmode \hbox{%
  \includegraphics[width=8.5cm]{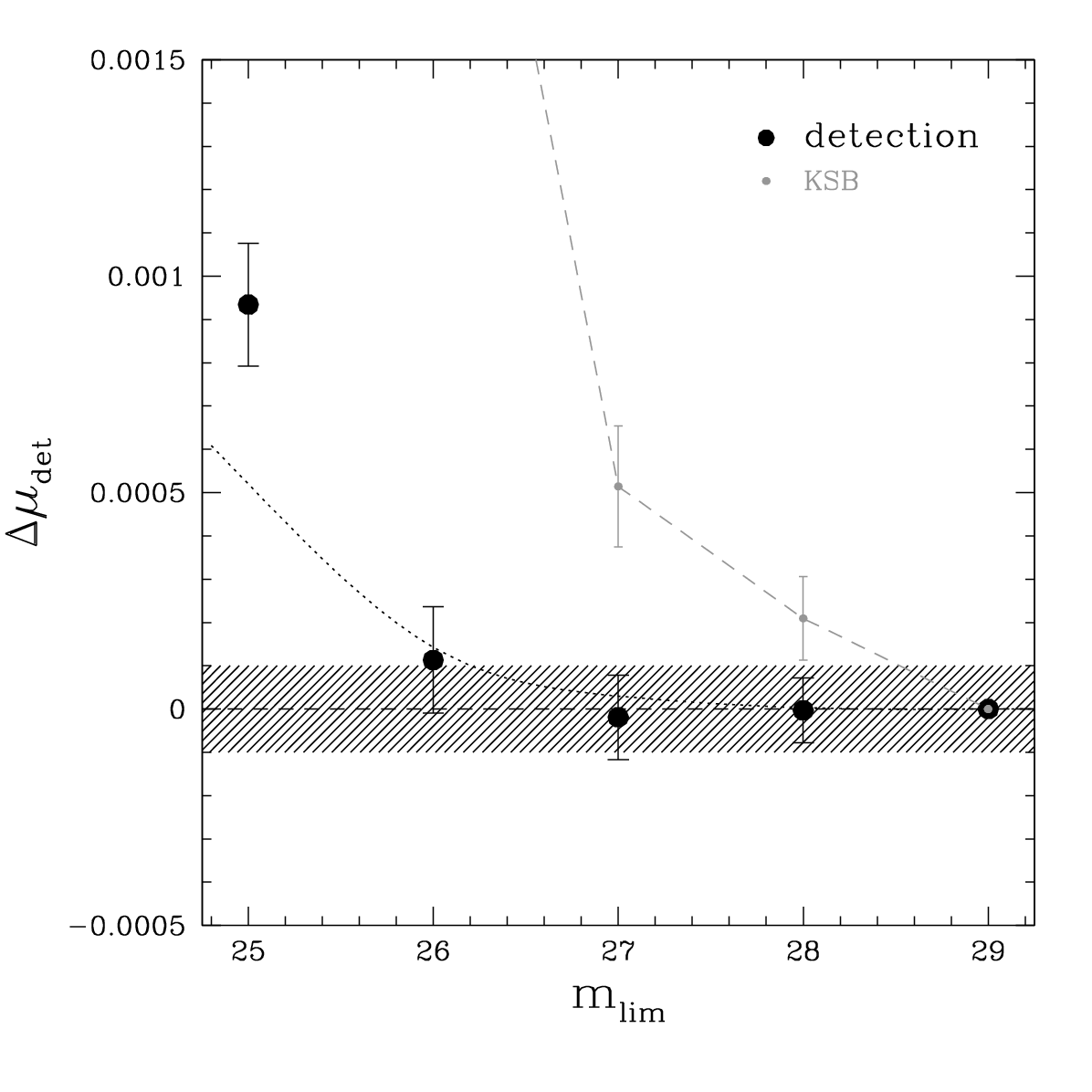}}
\caption{Change in multiplicative detection bias $\Delta\mu_{\rm det}$ (with respect to $\mu(m_{\rm lim}=29$)) for galaxies with $20<m_{\rm AUTO}<24.5$ as a function of $m_{\rm lim}$, the magnitude of the faintest galaxies that are included in the simulation (black points). The dotted line shows the change in bias if we select galaxies based on their input magnitude ($20<m_{\rm input}<24.5$). The change in multiplicative bias for the KSB algorithm is indicated by the light grey points. The hatched region indicates a tolerance of $10^{-4}$.
  \label{fig:detbias_maglim}}
\end{figure}

The blending of galaxies is a significant source of shear bias, and for a reliable estimate of the bias it is therefore critical to capture this in the simulated data. Studying the performance of galaxies on a grid may help in the comparison of methods, or the training of machine learning approaches \citep{Gruen10, Tewes19, Pujol20}, but the actual estimate relies on realistic simulations. In this Section we explore how the detection bias depends on the properties of the simulated galaxies, such as their size and ellipticity distributions. 

The realism is, however, not limited to the properties of the detected galaxies, because the performance of the shape measurements is also influenced by the presence of galaxies below the detection limit. This was first demonstrated  by \cite{Hoekstra15} for ground-based observations. Similarly, \citetalias{Hoekstra17} showed that for the \Euclid-like data we consider here, the multiplicative shear bias depends on $m_{\rm lim}$, the apparent magnitude of the faintest galaxies that are included in the image simulation. They found that galaxies as faint as $m_{\rm lim}=29$ can modify the multiplicative bias for the KSB algorithm.

The impact of very faint galaxies was studied in more detail in \cite{Martinet19} who found that the dependency with $m_{\rm lim}$ also depends on the shape measurement algorithm, and how it deals with blending. In our KSB setup the nearby objects are crudely masked, but no attempt is made to estimate the correct the surface brightness profile, thus biasing the estimates of the moments. Model fitting methods will generally do better in this regard, in line with the findings of \cite{Martinet19}. The clustering of galaxies results in a higher level of blending around brighter galaxies, and consequently, \cite{Martinet19} showed that the clustering of the faint galaxies increases the overall bias further. We do not consider this additional complication here, but note its importance when one aims to calibrate a shear measurement algorithm to be applied to actual data.

These studies only considered the final shear bias, but in Fig.~\ref{fig:detbias_maglim} we show how the \sextractor\ detection bias depends on $m_{\rm lim}$. Our results show that the detection bias is much less sensitive to the inclusion of faint galaxies, especially when compared to the KSB shear estimates (indicated by the light grey points and dashed line). The dotted line indicates the change in bias when we select galaxies based on $m_{\rm input}$. This shows that the bias partly arises from faint galaxies, for which the detection bias is larger (see Fig.~\ref{fig:detbias_mag}), scattering into the sample of sources used in the analysis (defined as $20<m_{\rm AUTO}<24.5$). Nonetheless, the convergence is only achieved for $m_{\rm lim}=27$, still $2.5$ magnitudes fainter than the magnitude limit of the sample of sources that we consider here. 

\begin{figure}
\centering
\leavevmode \hbox{%
\includegraphics[width=8.5cm]{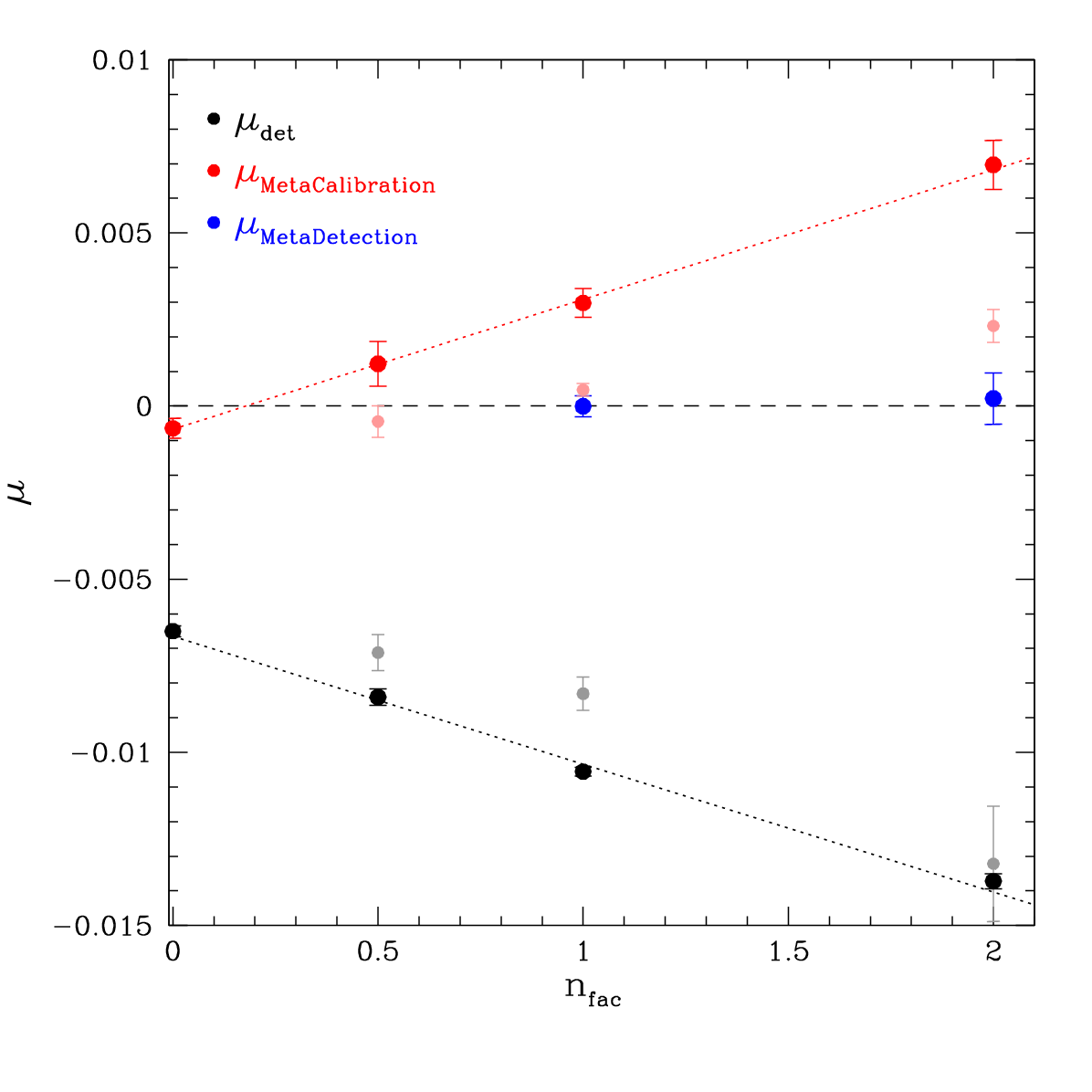}}
\caption{Multiplicative bias as a function of $n_{\rm fac}$, the relative increase in galaxy number density with respect to the baseline simulation. The grid-based results correspond to $n_{\rm fac}=0$. The black points show how the detection bias increases with $n_{\rm fac}$. The red (blue) points correspond to the \metacal\ (\metadet) results discussed in Sect.~\ref{sec:metacal} (Sect.~\ref{sec:meta_detect}). The light coloured points show the biases for relatively isolated galaxies (distance to nearest galaxy in the input catalogue larger than $2''$).
  \label{fig:nfac}}
\end{figure}

\subsection{Sensitivity to galaxy number density}
\label{sec:density}

\citetalias{Hoekstra17} (their Fig.~5) showed that the KSB shear bias increases if the number density of the simulated galaxies is increased by a factor $n_{\rm fac}$ (also see Table~\ref{tab:bias}). Consequently the bias will be larger near clusters and groups of galaxies, thus coupling the shear bias to the large-scale structure, which will need to be accounted for as shown by \cite{Hartlap11}.
An increase in detection bias will play a role, because Fig.~\ref{fig:detbias_separation} shows that it depends on the distance to the nearest galaxy. As the density increases, the mean separation decreases and the bias increases accordingly.

\begin{figure*}
\centering
\leavevmode \hbox{%
  \includegraphics[width=8.5cm]{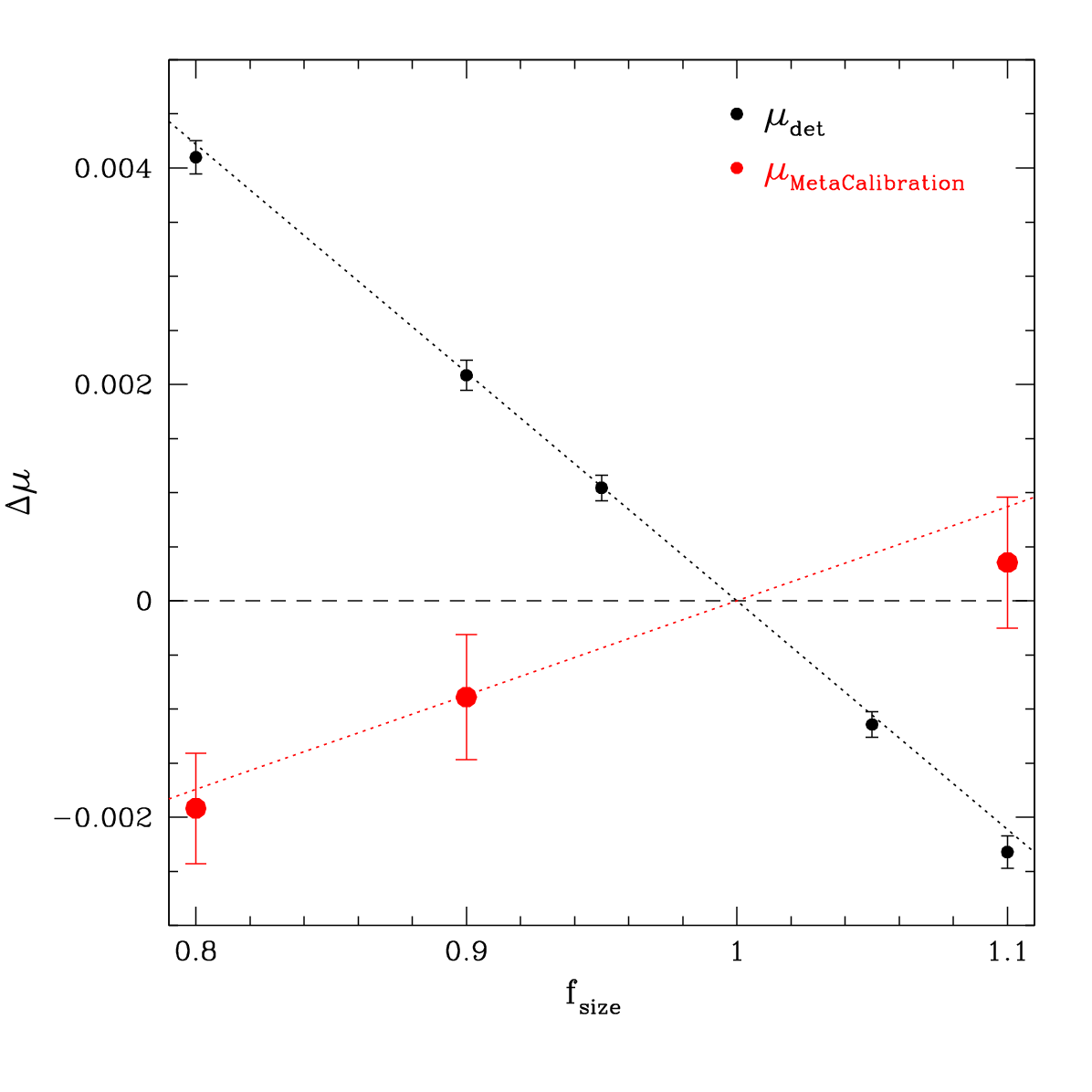}
  \includegraphics[width=8.5cm]{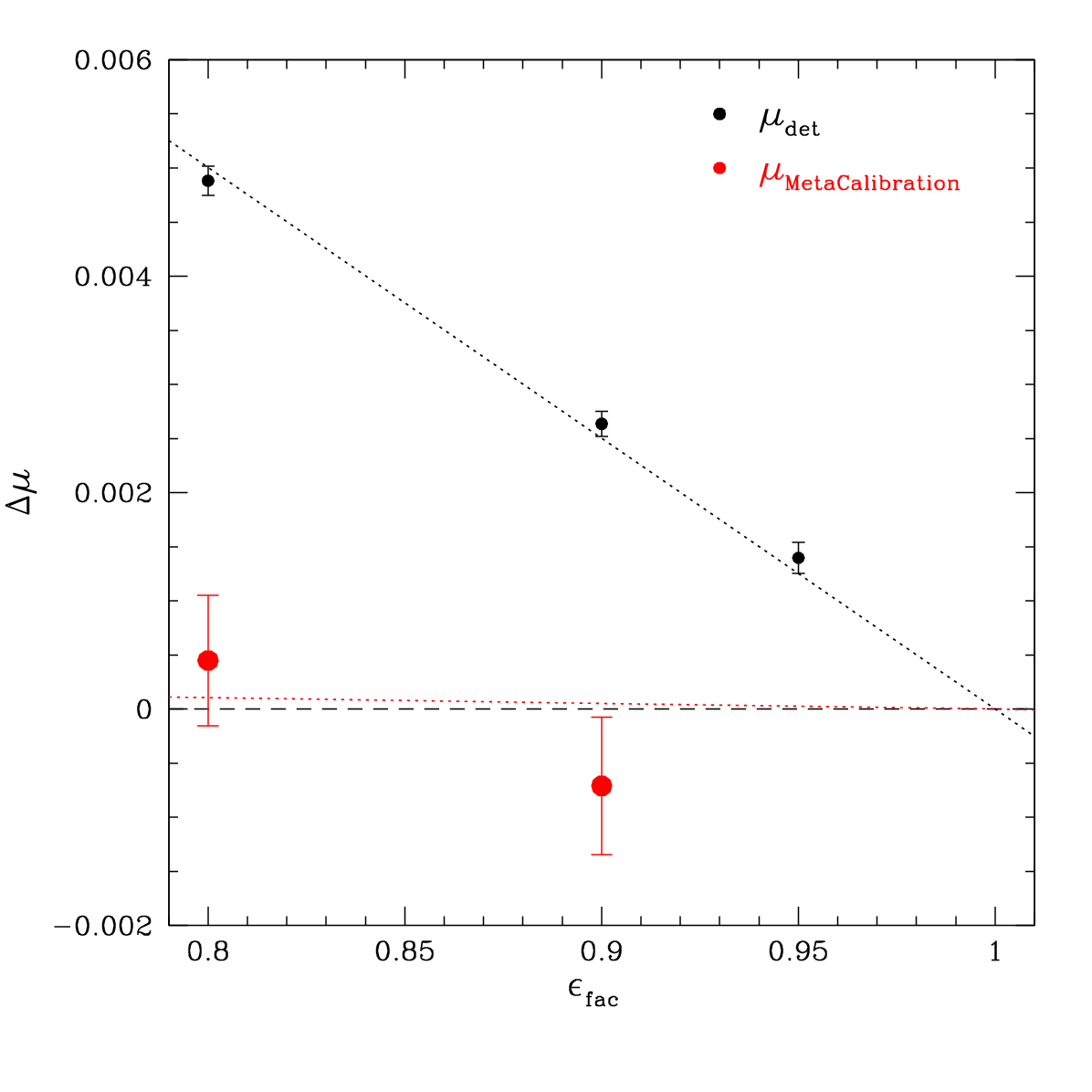}}
\caption{{\it Left panel:} change in multiplicative shear detection bias $\Delta\mu$ as a function of $f_{\rm size}$, the relative change in input galaxy size (black points).  {\it Right panel:} change in multiplicative detection bias if the input ellipticities are multiplied by a factor $\epsilon_{\rm fac}$. The dotted lines show the best fit linear model. The red points in both panels correspond to the post-\metacal\ results discussed in Sect.~\ref{sec:metacal}.
  \label{fig:morphology}}
\end{figure*}

We quantify the sensitivity of the detection bias to the galaxy number density in Fig.~\ref{fig:nfac}. The black points show $\mu_{\rm det}$ as a function of $n_{\rm fac}$, where $n_{\rm fac}=0$ corresponds to the grid-based simulations (no blending) and $n_{\rm fac}=1$ is our baseline case. For reference, a value of $n_{\rm fac}=2$ roughly corresponds to the galaxy density in the innermost regions of a massive cluster of galaxies \citep[see e.g. Fig.~11 in][]{Hoekstra15}. We find that the detection bias increases linearly with increasing galaxy density, with $\partial\mu_{\rm det}/\partial n_{\rm fac}=-0.003\,69\pm 0.000\,13$. As blending is a likely cause we repeat the measurements for a sample of relatively isolated galaxies (i.e., no neighbour brighter than $m=26$ in the input catalogue within $2''$ ) and show the results as light grey points in Fig.~\ref{fig:nfac}. The slope is almost halved, but not fully eliminated.

The spatial variation in $n_{\rm fac}$ caused by the clustering of galaxies will lead to spatial variations in the multiplicative bias across the survey. Provided these variations are small, the impact on the cosmological signal is expected to be negligible, as shown in \cite{Kitching19}. However, it is important that the galaxy number density in the simulations matches the average value in the survey, because a mismatch results in an overall shift in the shear bias. We discuss the area of high-quality data that is needed to achieve this in Appendix~\ref{app:galden}.

\subsection{Sensitivity to morphology}
\label{sec:morphology}

The detection bias depends on the morphology of the galaxies, because the size affects the signal-to-noise ratio and the incidence of blending. Moreover, the bias depends on the intrinsic ellipticity: the detection bias vanishes if $\epsilon^{\rm s}=0$, whereas we observe a significant detection bias for our reference setup. Such dependencies on morphology are of particular concern, because they vary with redshift \citep{Kannawadi15}, and can link shear biases to the lensing signal as the morphology depends on the galaxy density: early type galaxies are generally larger and rounder, and occupy higher density regions. Moreover, their photometric redshifts are typically more precise thanks to their more pronounced 4000\AA\ break, coupling the shear measurements to the binning of galaxies into tomographic bins. These connections highlight the need for simulations that capture the full process of photometric redshift and shear estimation simultaneously. This is, however, left for future study.

To explore the impact of uncertainties in the morphology further we analysed images where the input sizes are increased by a factor $f_{\rm size}$ and where the input ellipticities are increased by a factor $\epsilon_{\rm fac}$, similar to what was done in \citetalias{Hoekstra17} (see their Fig.~4 and~10). The black points in  Fig.~\ref{fig:morphology} show the change in bias as a function of these parameters. The left panel of Fig.~\ref{fig:morphology} shows that the detection bias increases linearly with increasing input galaxy sizes, with a slope $\partial\mu_{\rm det}/\partial f_{\rm size}=-0.0211\pm0.0006$. We expect the bias to be smaller if the galaxies are smaller, because the galaxies will be detected with a higher signal-to-noise ratio for a given magnitude, whilst blending is reduced. Although this dependence is rather steep, the distribution of galaxy sizes is fairly well established, and mismatches between the simulations and the data can be accounted for empirically \citepalias[see the discussion in][]{Hoekstra17}.

The sensitivity to the input ellipticity distribution is more worrisome, because it is generally more difficult to infer from existing high-quality {\it Hubble} Space Telescope observations. We find $\partial\mu_{\rm det}/\partial\epsilon_{\rm fac}=-0.025\,02\pm0.000\,56$, which is about half the value that \citetalias{Hoekstra17} measured for the full KSB bias. This suggests that a significant part of the sensitivity to the input ellipticity distribution is determined by the detection bias. Deeper observations may help improve empirical constraints on the ellipticity distribution \citep{Viola14}, but the measurements still require an accurate algorithm to measure shapes. Moreover, the results presented in Sect.~\ref{sec:noise} suggest that blending limits the gain of such deeper observations. This requires further study, because \cite{Kannawadi19} showed that the ellipticity distribution correlates with galaxy size and changes with redshift, whilst ellipticity gradients will complicate matters further. 

\begin{figure}
\centering
\leavevmode \hbox{%
  \includegraphics[width=8.5cm]{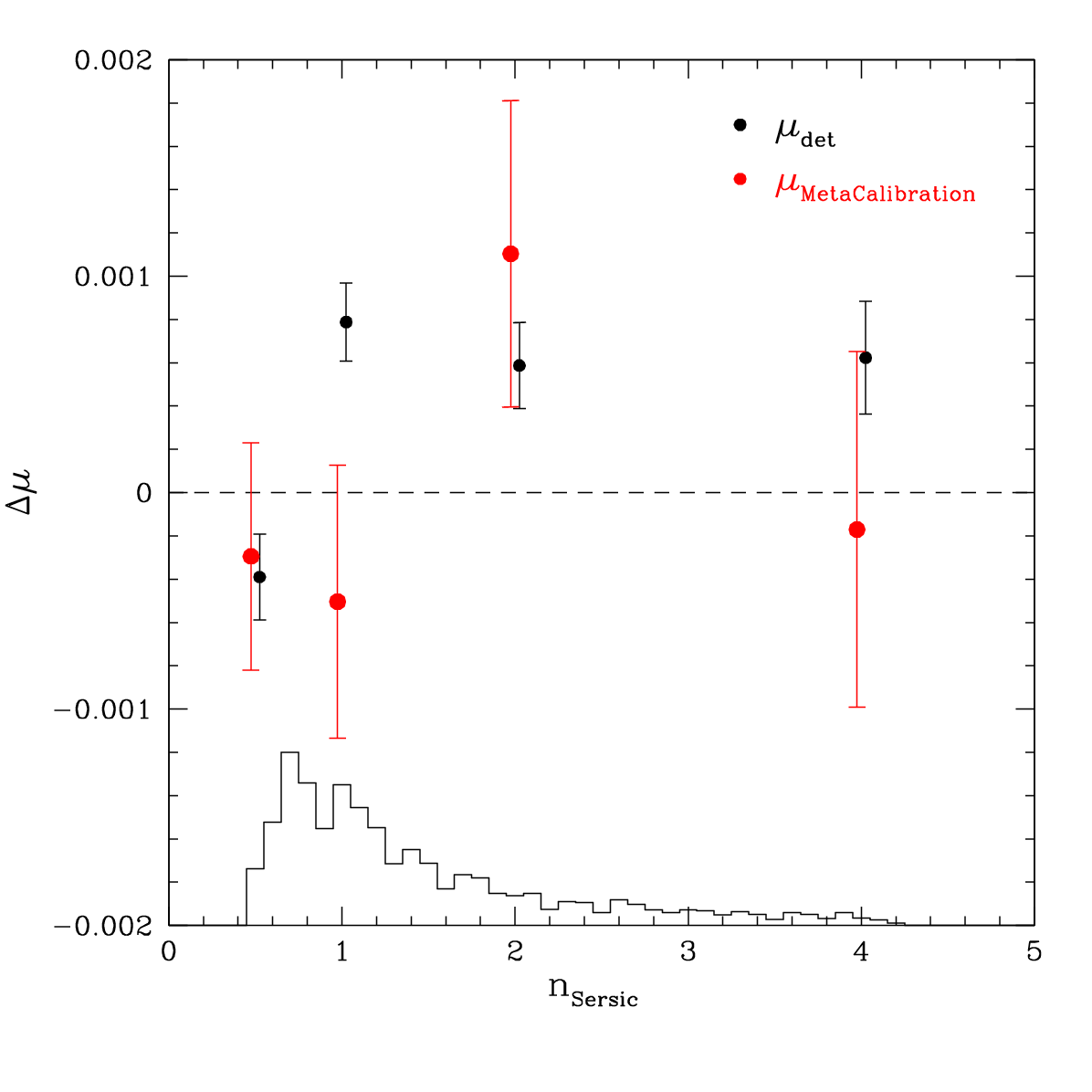}}
\caption{Change in multiplicative shear bias $\Delta\mu$ as a function \sersic\ index. The histogram indicates the distribution of \sersic\ indices in the baseline simulations. The black points show the change in detection bias. The red points show the \metacal\ results. 
\label{fig:bias_sersic}}
\end{figure}

The left panel of Fig.~\ref{fig:morphology} shows that the detection bias is reduced if galaxies are smaller, as such galaxies are easier to detect, whilst blending is reduced. We therefore expect the radial surface brightness profile to influence the bias as well. We explore two modifications, namely the sensitivity to changes in the \sersic\ index, $n_{\rm Sersic}$, and $r_{\rm trunc}$, the radius where the profile is truncated.

The black points in Fig.~\ref{fig:bias_sersic} show the change in detection bias when we keep the effective radii, fluxes and ellipticities of the galaxies the same, but fix the \sersic\ indices to a single value. Larger values for $n_{\rm Sersic}$ 
result in profiles that are more centrally peaked, reducing the detection bias.
Indeed, the bias slightly reduced with respect to the baseline case for $n_{\rm Sersic}\ge 1$. The histogram shows the baseline distribution of $n_{\rm Sersic}$, which peaks at values $<1$.

\begin{figure}
\centering
\leavevmode \hbox{%
  \includegraphics[width=8.5cm]{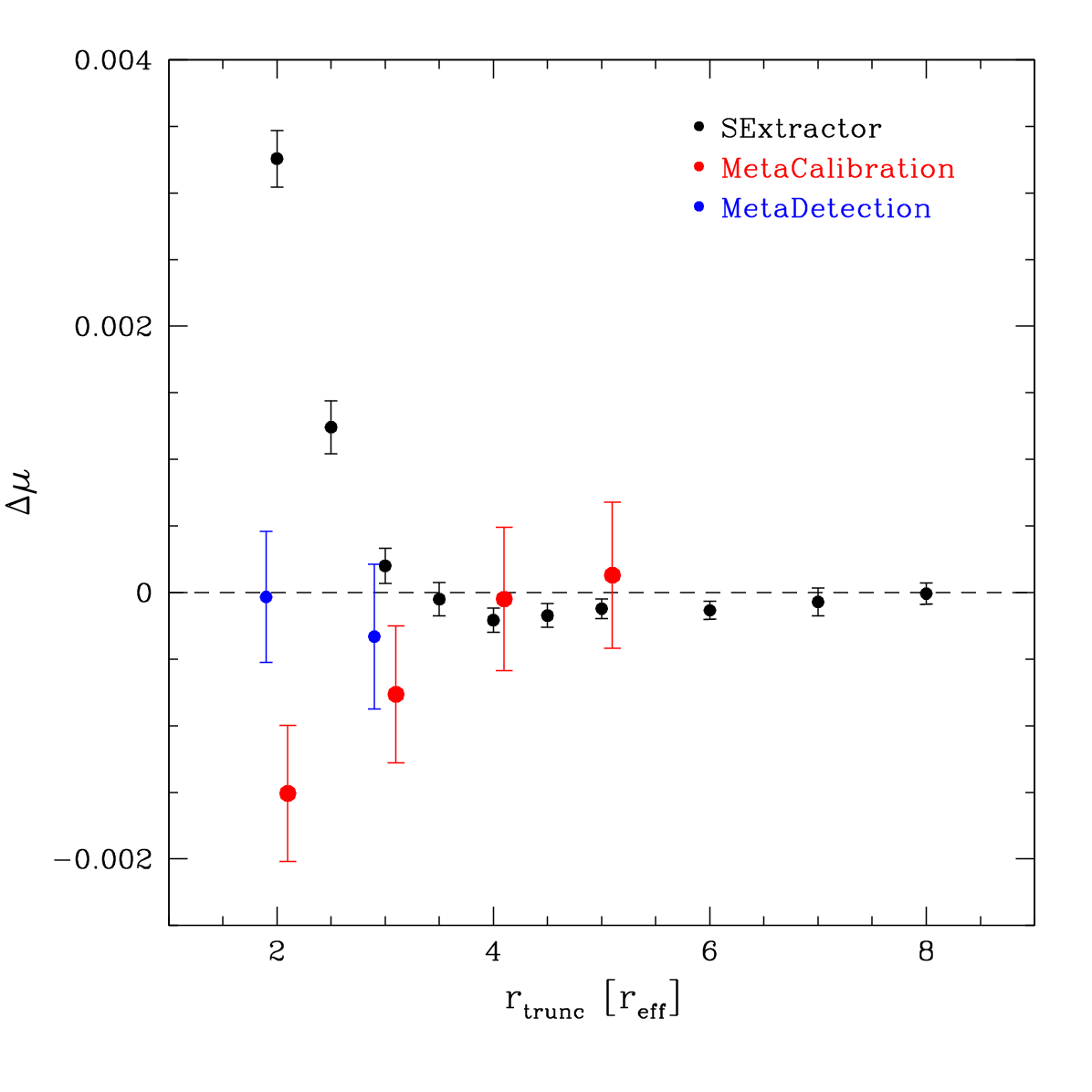}}
\caption{Change in multiplicative shear bias $\Delta\mu$ as a function of $r_{\rm trunc}$, the radius where the galaxy profile is truncated in the simulated images in units of the input half-light radius, $r_{\rm eff}$. The black points show the change in \sextractor\ detection bias. The red (blue) points show the \metacal\ (\metadet) results discussed in
Sect.~\ref{sec:metacal} (Sect.~\ref{sec:meta_detect}).
\label{fig:bias_galtrunc}}
\end{figure}

Throughout this paper we assume that the surface brightness profiles of galaxies are described by a \sersic-profile, which are truncated at 
$r_{\rm trunc}=3.5$ effective radii for reasons of computational speed. This ignores much of the variety in galaxy morphology, where the bulge and disk components may have different ellipticities and orientations. Moreover, spiral structure complicates matters further. Better modelling of the morphologies of galaxies using deep, high-quality data will help addressing this specific problem. A less explored question, however, is the surface brightness profile at large radii. \cite{Tal11} stacked the images of a large sample of luminous red galaxies and found that a \sersic-profile describes the data well out to more than 7 effective radii. In contrast, detailed studies of edge-on spiral galaxies indicate that the disks are truncated around 4 disk scale lengths on average \citep{Kregel02}.

We therefore created images where we truncate the profile at different values for $r_{\rm trunc}$ (in units of the effective radius $r_{\rm eff}$). The black points in Fig.~\ref{fig:bias_galtrunc} show the change in \sextractor\ detection bias, relative to the case of $r_{\rm trunc}=10$. The change is small for $r_{\rm trunc}>3.5$, indicating that it is important to accurately capture the surface brightness out to these radii.

The results presented in this section highlight the importance of capturing the morphological diversity of galaxies with sufficient accuracy. This seems quite feasible in the case of detection bias alone, but we expect the actual shear bias to be affected more. This is evidence from Fig.~\ref{fig:detbias_maglim}, where the KSB bias is sensitive to very faint galaxies, whereas the \sextractor\ detection bias converges at $m_{\rm lim}=27$ already. Similarly, \citetalias{Hoekstra17} found steeper dependencies for many parameters. The key question is therefore
whether image simulations can be made sufficiently realistic to capture the redshift-dependent morphologies of galaxies for Stage IV surveys. As this appears to be challenging, we explore next a different approach that uses the survey data to calibrate the shear estimate instead.

\section{MetaCalibration}\label{sec:metacal}

A different approach is to use the observations themselves to determine the response of an ensemble of galaxies to a shear. \cite{Huff17} worked out how one can estimate the shear bias by shearing the images, whilst taking the PSF and noise into account. They refer to this data-driven approach as \metacal\ and in this section we explore its potential to calibrate the multiplicative bias for our \Euclid-like simulations. In principle \metacal\ can also be used to correct for PSF anisotropy, but in the following we only consider the calibration of multiplicative shear bias, which allows us to limit the study to our round Airy PSF. 

The only assumption of \metacal\ is that we can construct a sheared version, $I^{\rm sh}({\bm x}|{\bm \gamma})$ of the true image using the observed image $I({\bm x})$ via Eq.~(5) of \cite{Huff17}):
\begin{equation}
I^{\rm sh}({\bm x}|{\bm \gamma})=P({\bm x})\ast [\hat s_{\gamma} \{P({\bm x})^{-1}\ast I({\bm x})\}].\label{eq:meta}
\end{equation}
where $\hat s_{\gamma}$ is the shear operator \citep{Bernstein02}, $P({\bm x})$ is the PSF, $I({\bm x})$ the observed image, and `$\ast$' indicates convolution. Hence, the observed image is first deconvolved ($P({\bm x})^{-1}\ast I({\bm x})$), then sheared by $\hat s_{\gamma}$, and finally re-convolved by the PSF. This procedure, in its simplest form, only requires an accurate model of the PSF. 

In practice, noise in the data complicates the deconvolution step, and a slightly larger PSF is needed to suppress the noise. The modified PSF, $P^{\rm meta}({\bm x})$, to use in the reconvolution step in Eq.~(\ref{eq:meta}) is \citep{Huff17}
\henk{
\begin{equation}
    P^{\rm meta}({\bm x})=P({\bm x}/(1+2|\gamma|)).
\end{equation}}

These steps implicitly assume that the images are well sampled, so that the image manipulations are not compromised. However, in the case of both \Euclid\ 
and the {\it Roman} Space Telescope, the pixels are large compared to the PSF size. In our calculations we do assume that we can construct a well-sampled model of the PSF, but the images of the smallest galaxies might still be affected by undersampling. In a companion paper, \cite{Kannawadi20} explore ways to mitigate this, but we note that image simulations can also be used to correct for the biases that may be introduced.

Another complication is that the shearing of the images leads to anisotropic correlated noise, which needs to be accounted for. One possibility is to determine the resulting bias using image simulations, but \citet[][SH17 hereafter]{Sheldon17} show that this problem can also be mitigated by adding anisotropic noise. The latter approach does lead to a slight increase in the overall noise level, but as shape noise typically dominates, this is only a minor concern. There are other complications that are particularly relevant for space-based observations, such as the wavelength-dependence of the PSF, which we discuss in more detail in Sect.~\ref{sec:discussion}. 

If we use Eq.~(\ref{eq:meta}) to apply a small shear ${\bm \gamma}=(\gamma_1,\gamma_2)$ to a galaxy image, and measure its shape ${\bm e}=(e_1,e_2)$ we can relate the resulting shape to the original value ${\bm e}_{{\bm \gamma}={\bm 0}}$, because
\begin{equation}
    {\bm e}\approx \bm{e}\rvert_{{\bm\gamma}={\bm 0}}+\frac{\partial {\bm e}}{\partial {\bm \gamma}}\biggr\rvert_{{\bm \gamma}={\bm 0}}{\bm \gamma}\equiv
    {\bm e}\rvert_{{\bm \gamma}=\bm{0}}+\bm{\mathsf{R}}^\gamma\,{\bm \gamma},
    \label{eq:sheartensor}
\end{equation}
where $\bm{\mathsf{R}}^\gamma$ is the $2\times 2$ shear response tensor. We can estimate its elements by measuring the shapes of the galaxies in the sheared images and computing
\begin{equation}
    {\mathsf R}^\gamma_{ij}=\frac{e^+_i-e^-_i}{\Delta\gamma_j}
\end{equation}
where the subscripts indicate the two shear components, and the superscript the sign of the applied shear, i.e. `$+$' means the image was sheared by $+\gamma_j$, etc; hence, $\Delta\gamma_j=2\gamma_j$. 

This expression is true for any shape measurement, and it allows us to estimate the shear, $\hat{{\bm \gamma}}$ for an ensemble of galaxies (as $\langle{\bm e}\rangle|_{{\bm \gamma}={\bm 0}}\approx {\bm 0}$)
\begin{equation}
\hat{{\bm \gamma}}\approx \langle\bm{\mathsf{R}}^\gamma\rangle^{-1}
\langle{\bm e}\rangle= \langle\bm{\mathsf{R}}^\gamma\rangle^{-1}
\langle\bm{\mathsf{R}}^\gamma\,{\bm \gamma}\rangle,\label{eq:shearmeta}
\end{equation}
where the shape measurements are obtained from the image that is
convolved with $P^{\rm meta}(\bm{x})$. Hence, we average the estimates
for the shapes and the shear responses, rather than using estimates
per galaxy. The reason is that the estimates for $\bm{\mathsf{R}}^\gamma$ are very noisy for individual galaxies, and averaging reduces biases in the shear estimate, which requires the inverse of $\bm{\mathsf{R}}^\gamma$. To reduce the noise even further we average the estimates for $\bm{\mathsf{R}}^\gamma$ for a particular selection of galaxies over many images (typically 3300). We verified that $\bm{\mathsf{R}}^\gamma$ does not change as a function of shear in the simulated images.  Moreover, we find that the off-diagonal elements vanish and we therefore assume that $\bm{\mathsf{R}}^\gamma$ is diagonal in the remainder of this paper. 

Equation~(\ref{eq:shearmeta}) shows that the resulting shear estimate for the ensemble of galaxies is actually weighted by $\bm{\mathsf{R}}^\gamma$, and hence one would like to use a shape measurement algorithm so that $\bm{\mathsf{R}}^\gamma\approx \bm{\mathsf{I}}$. This is, however, not an immediate concern for our study, because the PSF is isotropic and the same shear is applied to all the simulated galaxies. 

In principle it should not matter what shape measurement we use, because any intrinsic bias in the estimator will be accounted for by \metacal. We therefore 
simply use the polarisation ${\bm \chi}$, 
\begin{equation}
    \chi_1=\frac{Q_{11}-Q_{22}}{Q_{11}+Q_{22}},\ {\rm and}\;
    \chi_2=\frac{2\,Q_{12}}{Q_{11}+Q_{22}},
\end{equation}
where the weighted quadrupole moments $Q_{ij}$ are defined as
\begin{equation}
     Q_{ij} = \int\,{\rm d}^2\mathbf{x}\,x_i x_j W(\mathbf{x})\, I(\mathbf{x}), 
\end{equation}
and $W(\mathbf{x})$ is the weight function, for which we use a Gaussian with a fixed value for the dispersion of $\sigma_{\rm w}$, i.e. we do not try to optimise the width of the weight function to each object, nor do we try to correct for blending of objects.

The use of a fixed value for $\sigma_{\rm w}$ has the advantage that the measurement does not depend on the observed size of the object, which will differ for the different sheared versions of the images as it correlates with the shear,
so that $\bm{\mathsf{R}}^\gamma$ fully captures the shear response in the absence of detection bias. We adopt $\sigma_{\rm w}=2$ pixels (i.e. $\sigma_{\rm w}=0\farcs2$) as our baseline, which is a reasonable value to use for the galaxies in our simulations, as suggested by Fig.~\ref{fig:detbias_filter}. Moreover, as shown in our companion paper \citep{Kannawadi20}, this weight function is wide enough to avoid aliasing bias.

We describe and test our \metacal\ setup using the grid-based simulations in Sect.~\ref{sec:meta_grid}. We study the performance on our more realistic baseline simulations in Sect.~\ref{sec:meta_sim}, which enable us to quantify the impact of blending. We also explore the sensitivity of the post-\metacal\ bias to changes in the galaxy number density and morphology. The prospects of \metadet\ \citep{Sheldon19} are examined in Sect.~\ref{sec:meta_detect}. 

\subsection{Grid-based simulations}
\label{sec:meta_grid}

\begin{figure}
\centering
\leavevmode \hbox{%
  \includegraphics[width=8.5cm]{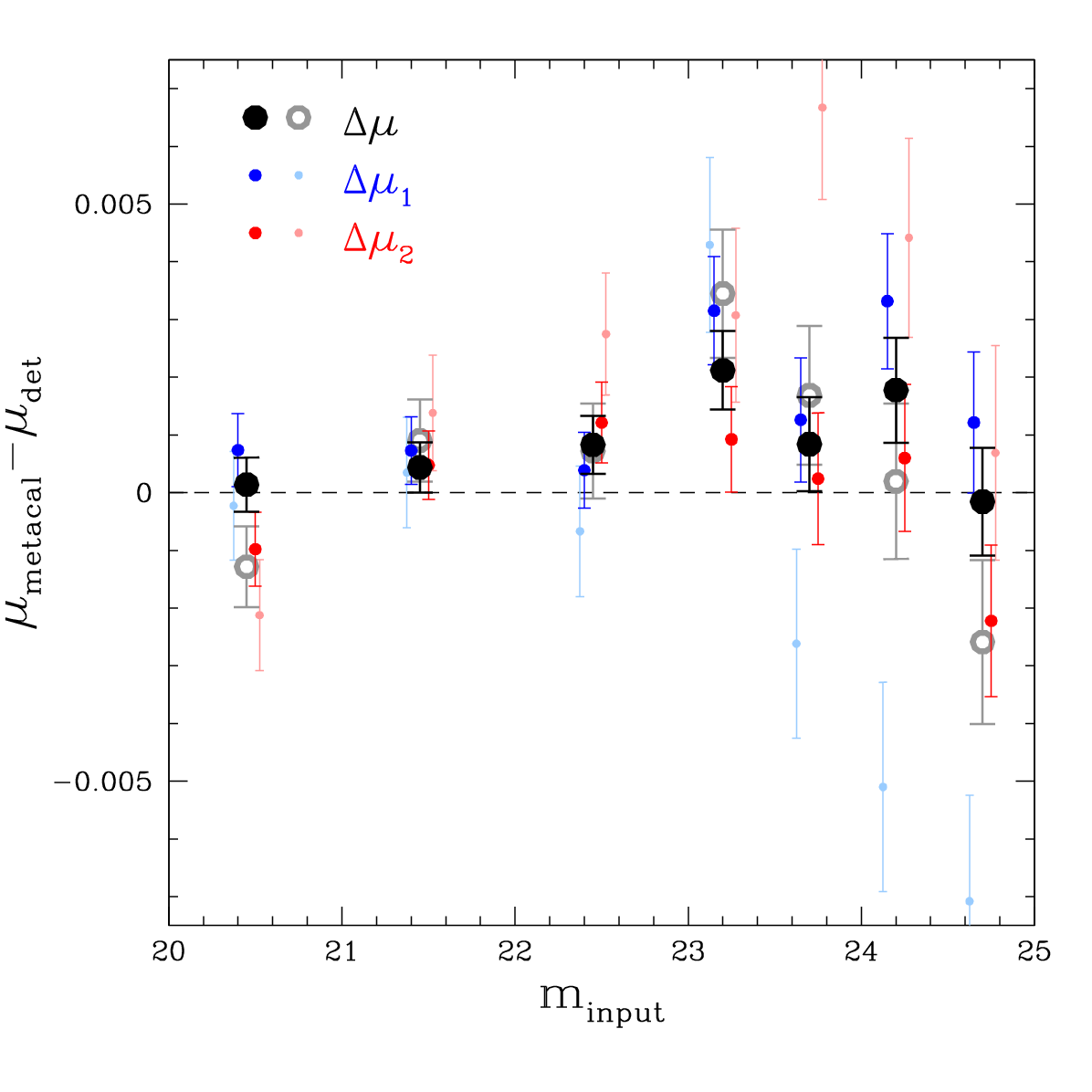}}
\caption{Difference between the multiplicative bias after \metacal, $\mu_{\rm metacal}$ and detection bias, $\mu_{\rm det}$, as a function of the input magnitude $m_{\rm input}$ for the grid-based simulations. The bright (light) colours show the results when we apply a shear of $\pm0.02$ ($\pm0.01$) in the metacalibration step. The solid black (open grey) points show the average bias, and the blue (red) points indicate $\Delta\mu_1$ ($\Delta\mu_2$). Using a larger shear results in smaller uncertainties and a better agreement between the two shear components. \label{fig:delta_mu_meta}}
\end{figure}

\citetalias{Sheldon17} presented a practical implementation of \metacal\footnote{\url{https://github.com/esheldon/ngmix}}, and we use the default setup here. Although the image manipulations can be done on postage stamps, we instead process the full simulated images. This naturally allows us to quantify selection biases as described in \citetalias{Sheldon17} and \cite{Sheldon19}. However, in this section we ignore the impact of selection bias.

We use the \metacal\ implementation in \galsim\ to create the five images needed to compute the shear response for the grid-based simulated images. To do so, we have to choose the value of $\Delta\gamma$ to use.  Applying a larger shear has the benefit of increasing the precision with which the shear response can be measured, but if the value is too large, higher order terms may become relevant. This was explored in \citetalias{Sheldon17} who found that for $\Delta\gamma<0.04$ the changes are negligible. We therefore consider two values for $\Delta\gamma$,
namely 0.02 and 0.04 and match the resulting shape measurements to the \sextractor\ catalogue.

As reported in Table~\ref{tab:bias}  we observe a significant multiplicative bias for both shear components, which agree with each other. If \metacal\ yields an unbiased shear estimate, the measured multiplicative bias, $\mu_{\rm meta}$, however, should recover the \sextractor\ detection bias, $\mu_{\rm det}$. Indeed, we find that that the bias that can be attributed to the shape measurements part is much smaller, with $\mu_1^{\rm meta}-\mu_1^{\rm det}=0.001\,65\pm 0.000\,46$ and $\mu_2^{\rm meta}-\mu_2^{\rm det}=-0.000\,86\pm 0.000\,43$ for galaxies with $20<m_{\rm AUTO}<24.5$, comparable to the requirements derived in \cite{Cropper13}. 

To explore the performance of \metacal\ further, we show $\mu_{\rm meta}-\mu_{\rm det}$ as a function of $m_{\rm input}$ in Fig.~\ref{fig:delta_mu_meta}. The use of the input magnitude ensures efficient shape noise cancellation. Comparison of the black ($\Delta\gamma=0.04$) and open grey ($\Delta\gamma=0.02$) points shows that the overall performance is similar, but that using a larger shear does indeed result in smaller uncertainties. Importantly, when we consider the  two shear components separately, we find that they differ for $\Delta\gamma=0.02$ (light coloured points) when $m_{\rm input}>23.5$, whereas  $\Delta\gamma=0.04$ (bright points) yields consistent values for $\mu_1$ and $\mu_2$. Sampling may play a role here \citep[see e.g.][]{Kannawadi20}, but as the differences vanish when we apply the larger shear, we  adopt this as our baseline. 

\begin{table*}
  \caption{Average biases after \metacal\ for galaxies with $20<m_{\rm AUTO}<24.5$.\label{tab:metacal_bias}}
  \centering
\begin{tabular}{lrrrr}
  \hline
  \hline
   setup &  $\mu_1$ & $\mu_2$  & $c_1~[\times 10^5]$ & $c_2~[\times 10^5]$ \\
   \hline
   \multicolumn{5}{c}{\bf baseline $\sigma_{\rm w}=2$ pixels}\\
   \hline
   \metacal\ ($\bm{\mathsf{R}}^\gamma$ only) & $-0.009\,35\pm 0.000\, 42$ & $-0.008\,80\pm 0.000\,38$ & $-0.73\pm 1.60$ & $-1.24\pm1.45$ \\
   \metacal\ (full) & $0.002\,74\pm 0.000\,42$ & $0.002\,89\pm 0.000\,38$ & $-0.74\pm1.62$ & $-1.26\pm1.47$ \\
   \metacal\ (full, $r_{\rm sep,in}>2''$) & $0.000\,00\pm 0.000\,25$ & $0.000\,95\pm 0.000\,26$ & $-0.18\pm0.96$ & $0.93\pm0.98$ \\
   \metacal\ (full, $r_{\rm sep,det}>2''$) & $-0.006\,94\pm0.000\,31$ & $-0.006\,01\pm 0.000\,30$ &
   $-1.38\pm1.18$ & $0.74\pm 1.14$\\
   \metadet\ & $-0.000\,81\pm 0.000\,40$ & $0.000\, 37\pm 0.000\, 41$ & $-2.90\pm 1.54$ & $0.43\pm 1.57$ \\
   \hline
   \multicolumn{5}{c}{\bf baseline $\sigma_{\rm w}=3$ pixels}\\ 
   \hline
   \metacal\ ($\bm{\mathsf{R}}^\gamma$ only) & $-0.012\,59\pm 0.000\,42$ & $-0.012\,23\pm0.000\,37$ & $-3.01\pm 1.60$ & $0.14\pm 1.40$ \\
   \metacal\ (full) & $0.001\,66\pm 0.000\,42$ & $0.001\,64\pm 0.000\,37$ & $-3.05\pm1.62$ & $0.14\pm1.42$ \\
   \metacal\ (full, $r_{\rm sep}^{\rm in}>2''$) & $0.000\,21 \pm 0.000\,27$ & $0.002\,37\pm 0.000\,25$ & $-0.96\pm1.05$ & $1.19\pm 0.96$ \\
   \metacal\ (full, $r_{\rm sep}^{\rm det}>2''$) & $-0.007\,14\pm 0.000\,29$ & $-0.005\,25\pm 0.000\,27$ & $-2.16\pm 1.11$ & $1.02\pm 1.03$ \\
     \hline
   \multicolumn{5}{c}{\bf grid $\sigma_{\rm w}=2$ pixels} \\
   \hline
   \metacal\ ($\bm{\mathsf{R}}^\gamma$ only) & $-0.008\,52\pm 0.000\,41$ & $-0.006\,99\pm 0.000\,37$ & $0.33\pm 1.57$ & $1.31\pm 1.41$ \\
   \metacal\ (full) & $-0.001\,07\pm0.000\,41$ & $-0.000\,36\pm 0.000\,37$ & $0.33\pm 1.58$ & $1.31\pm 1.42$ \\
  \hline
  \hline  
\end{tabular}
\bigskip
\begin{minipage}{\linewidth}
\tablefoot{In the `baseline' case the galaxies are placed randomly and their images are sheared. Galaxies are placed on a regular grid, about $9''$ apart for the `grid' results. See text for details on the various selections.}
\end{minipage}
\end{table*}

\subsection{Baseline results}
\label{sec:meta_sim}

We now proceed to use the setup with $\Delta\gamma=0.04$ and $\sigma_{\rm w}=2$ pixels to examine the performance of \metacal\ on the simulations where galaxies are positioned randomly. Moreover, we explore the possibility to account for the selection bias using the procedure outlined in \citetalias{Sheldon17}. Although our use of a fixed weight function avoids introducing a weight bias\footnote{The size estimated from the best-fit Gaussian is different for the two image rotations after a shear has been applied. Using the observed size would thus couple the weight function to the shear itself, leading to a bias.}, the selection bias introduced by \sextractor\ remains.

\citetalias{Sheldon17} show how the selection bias can be included in \metacal, by noting it introduces an ellipticity\footnote{We use ellipticity here as a synonym for shape, but note that the discussion is independent of the estimator employed.} dependent weighting, $S({\bm e})$, of an underlying ellipticity distribution $P({\bm e})$. Hence the ensemble averaged mean ellipticity can be expressed as
\begin{equation}
    \langle {\bm e}\rangle^{\rm S}=\int {\rm d}{\bm e}\, S({\bm e})\,P({\bm e})\,{\bm e},
\end{equation}
where we assume that $\int d{\bm e}S({\bm e})P({\bm e})=1$. We can express the ensemble averaged version of Eq.~(\ref{eq:sheartensor}) as
\begin{align}
    \langle\bm{\mathsf{R}}\rangle & =\int {\rm d}{\bm e}\,\frac{\partial [S({\bm e})\,P({\bm e})\,{\bm e}]}{\partial{\bm \gamma}}\biggr\rvert_{{\bm\gamma}={\bm 0}}\nonumber\\
 &=\int {\rm d}{\bm e}\,\left[ S({\bm e})\, \frac{\partial [P({\bm e})\, {\bm e}]}{\partial {\bm \gamma}}\bigg|_{{\bm \gamma}={\bm 0}} +  P({\bm e})\,{\bm e}\,\frac{\partial S({\bm e})}{\partial {\bm \gamma}}\bigg|_{{\bm \gamma}=\bm{0}} \right]\nonumber\\
 & \equiv \langle\bm{\mathsf{R}}^{\gamma}\rangle + \langle\bm{\mathsf{R}}^{\rm S}\rangle.\label{eq:metafull}
\end{align}

If there is no selection bias, i.e. $S({\bm e})=1$, the second term in Eq.~(\ref{eq:metafull}) vanishes and we can identify the first term with $\bm{\mathsf{R}}^\gamma$. The second term quantifies the response of the shear estimate to the selection bias. As discussed in \citetalias{Sheldon17}, $\bm{\mathsf{R}}^{\rm S}$ can be estimated by measuring the mean ellipticity from the unsheared image, but selecting the measurements from the sheared images. The \metacal\ estimate of the selection bias is then
\begin{equation}
    \mu^{\rm sel}_i=\frac{\mathsf{R}^\gamma_{ii}+\mathsf{R}^{\rm S}_{ii}}{\mathsf{R}^\gamma_{ii}}.
\end{equation}

\begin{figure*}
\centering
\leavevmode \hbox{%
  \includegraphics[width=8.5cm]{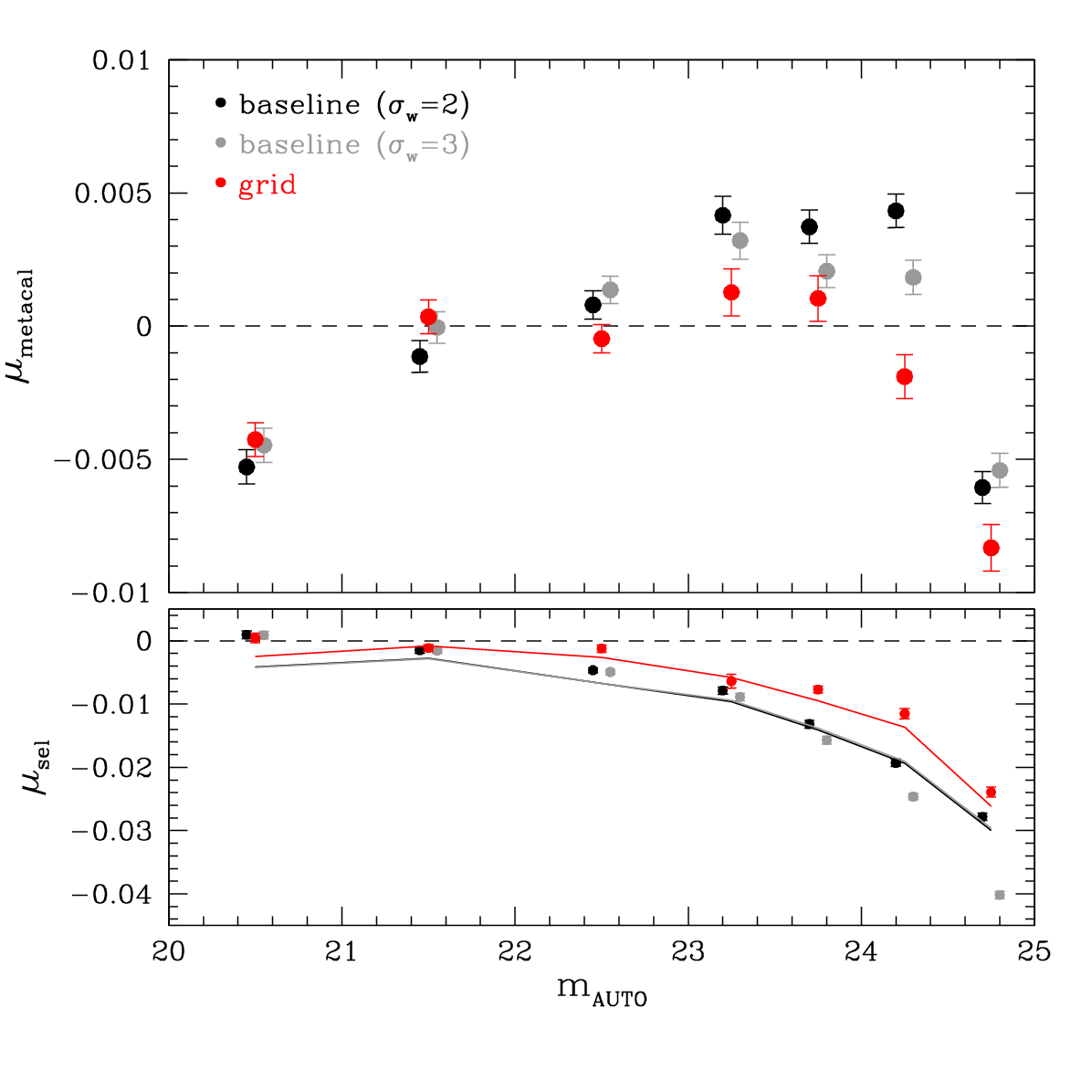}
   \includegraphics[width=8.5cm]{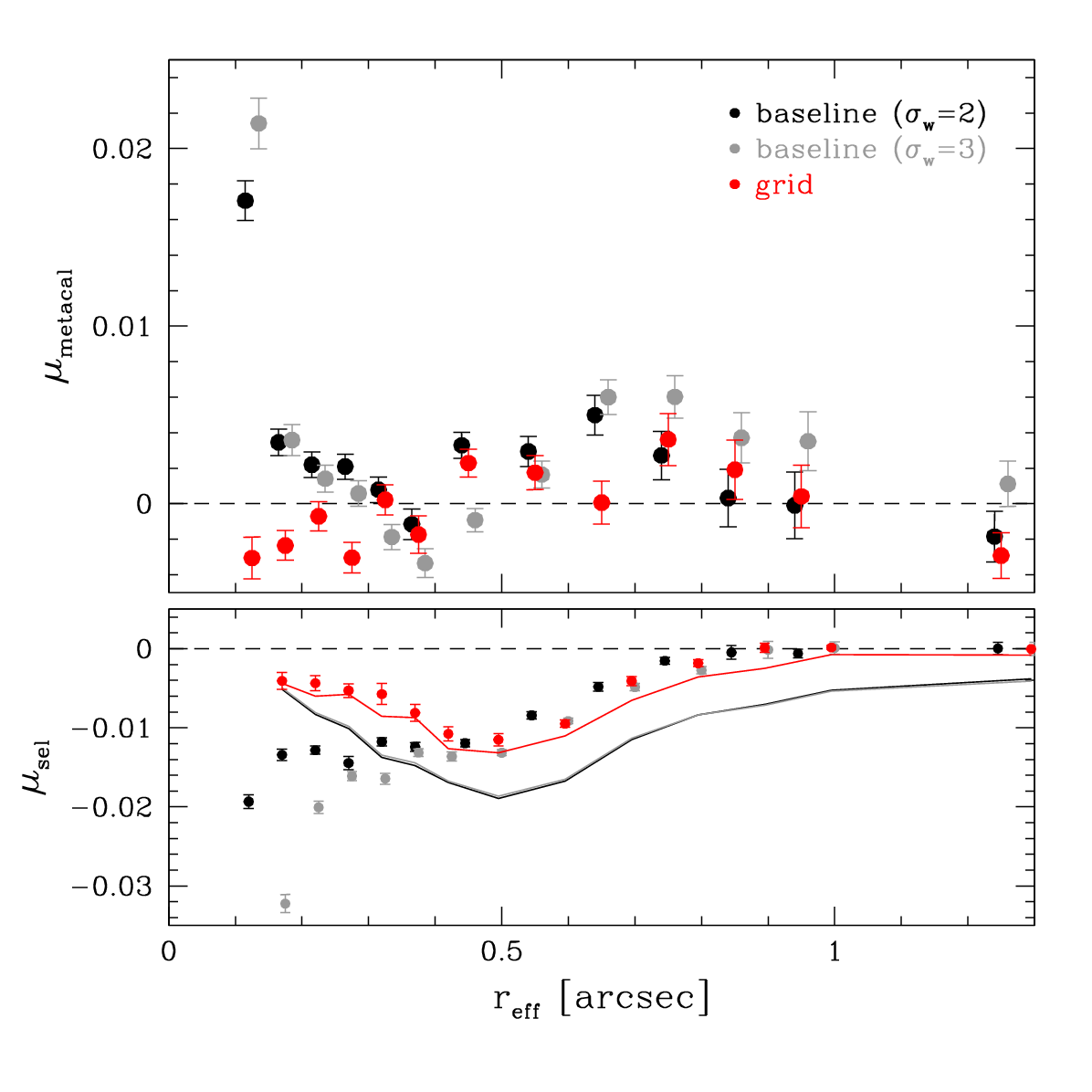}}
\caption{{\it Left panel}: Multiplicative bias after full \metacal\ as a function of $m_{\rm AUTO}$ for the baseline simulations 
(black for $\sigma_{\rm w}=2$ pixels; lightgrey for $\sigma_{\rm w}=3$ pixels) and the grid-based simulations (red points for $\sigma_{\rm w}=2$ pixels). {\it Right panel}: Multiplicative bias after full \metacal\ as a function of the input half-light radius ($r_{\rm eff}$) for galaxies with $m_{\rm AUTO}$. The bottom panels show the estimated selection bias from full \metacal\ (points). The solid lines show the corresponding direct measurements of the selection bias \henk{(c.f.~Figs.~\ref{fig:detbias_mag} and~\ref{fig:bias_size}).}
  \label{fig:metacal_full}}
\end{figure*}

The left panel in Fig~\ref{fig:metacal_full} shows the resulting multiplicative bias after \metacal\ when we account for the selection bias as a function of the observed apparent magnitude. The residuals for the grid-based results (red points) are very small, except for the galaxies with $m_{\rm AUTO}>24.5$. The bottom panel shows the \metacal\ estimates for the selection bias, which agrees well with the actual bias that we infer from comparison to the input catalogue.

We report the mean biases for the two shear components in Table~\ref{tab:metacal_bias} for galaxies with $20<m_{\rm AUTO}<24.5$. For the grid-based simulations we find $\langle\mu\rangle=0.000\,64\pm 0.000\, 29$, i.e. well within requirements for Stage IV surveys. For the baseline case (black points) the results are similar, but we do observe a significant residual bias $\langle\mu\rangle=0.002\,88\pm 0.000\, 29$, driven by galaxies with $m_{\rm AUTO}>23$. For reference we also repeated the measurements using a wider weight function with $\sigma_{\rm w}$=3 pixels, and we obtain similar results (see Table~\ref{tab:metacal_bias}).

The right panel of Fig.~\ref{fig:metacal_full} shows the post-\metacal\ bias as a function of the input galaxy size. For the grid based simulations (red points) the bias is flat as a function of size, thus effectively correcting for the detection bias (shown in the bottom panel, as well as Fig.~\ref{fig:bias_size}. The biases are also small for the baseline case, with both weight functions yielding consistent results. Only for the smallest galaxies do we observe a significant bias, which is not seen when galaxies are placed on a grid. This rules out sampling as the cause, but rather points to blending. Indeed, if we limit the comparison to isolated galaxies (no neighbour within $2''$ in the input catalogue with $m<26$), the results are similar to the grid-based simulations.

\begin{figure*}
\centering
\leavevmode \hbox{%
  \includegraphics[width=8.5cm]{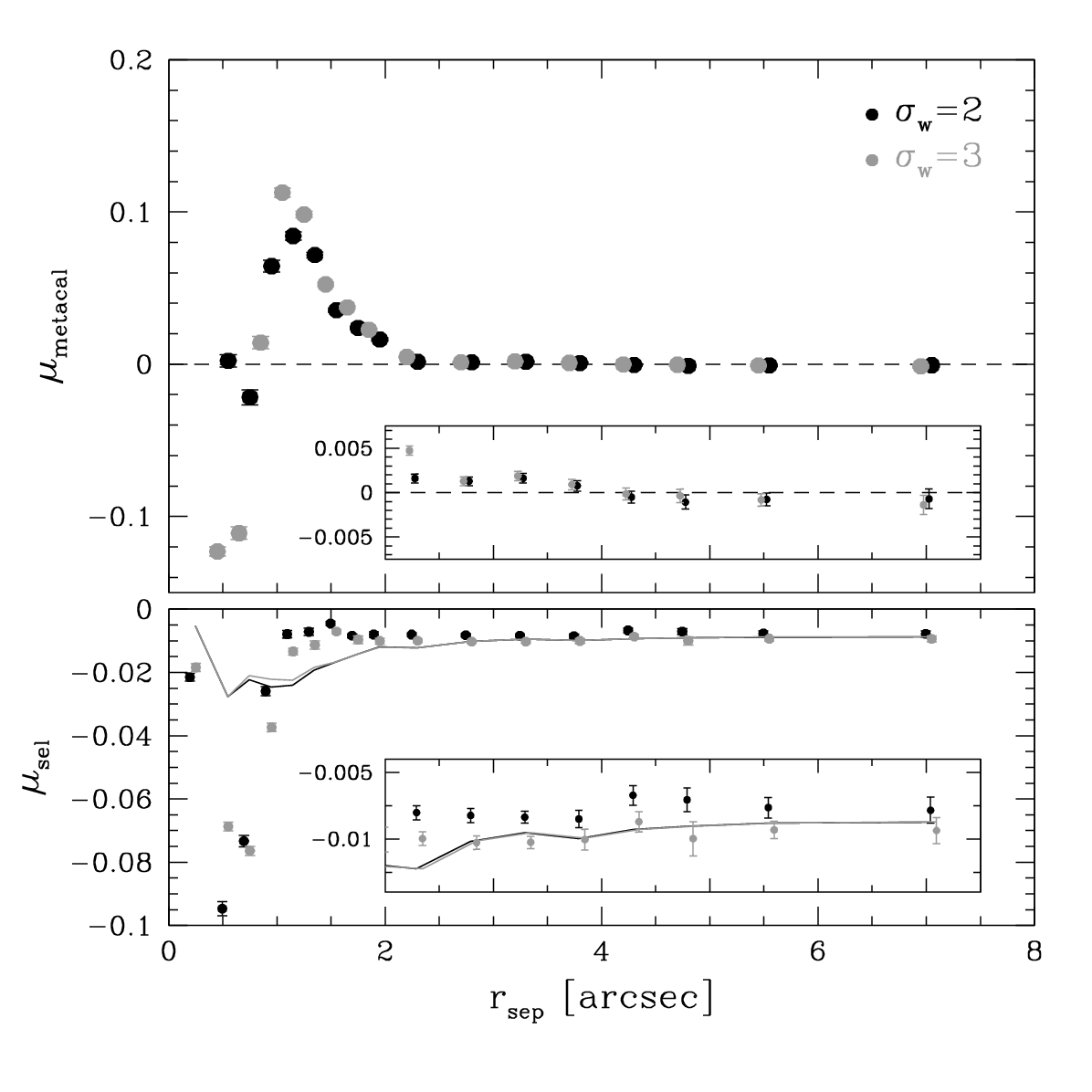}
  \includegraphics[width=8.5cm]{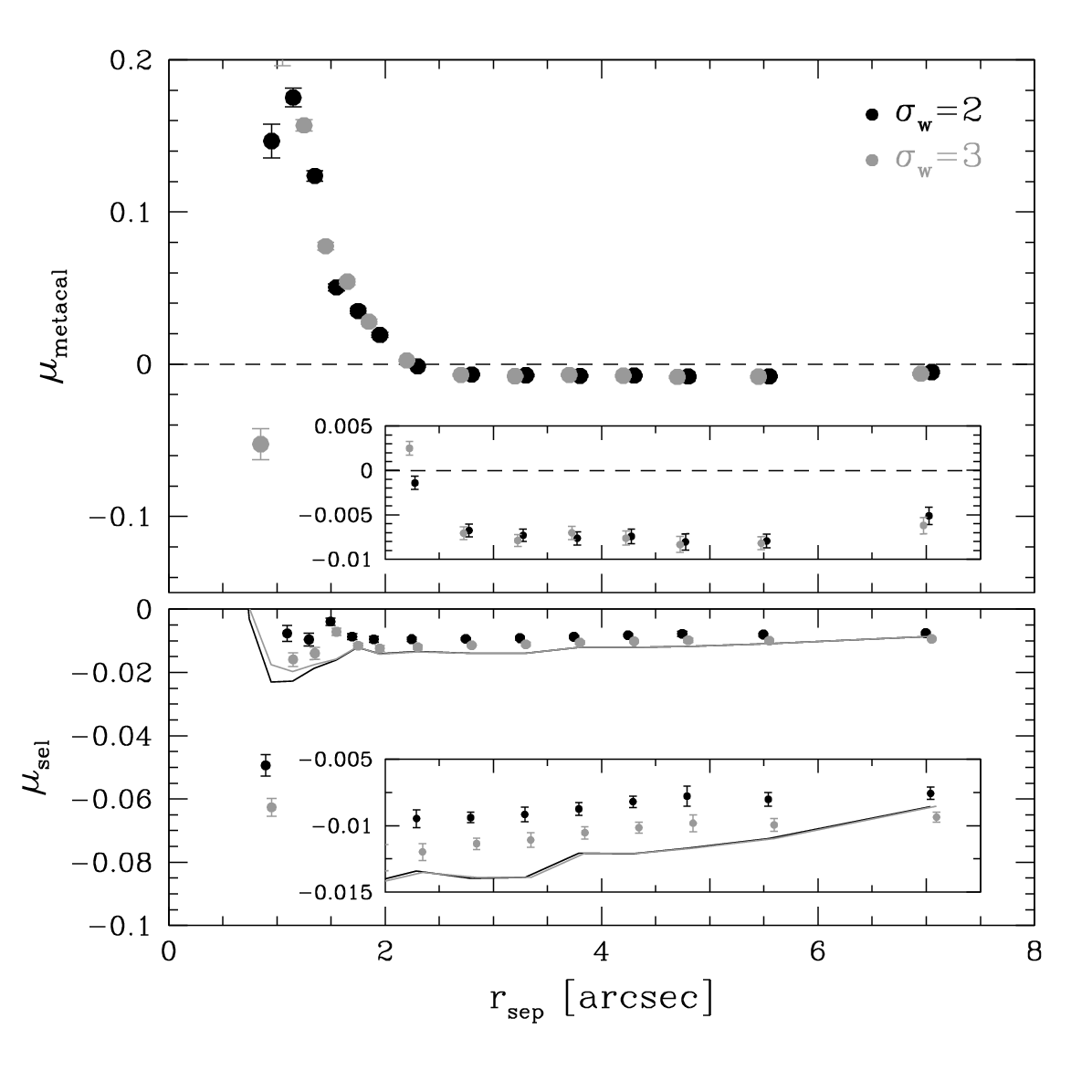}}
\caption{{\it Left panel}: Multiplicative bias after full \metacal\ for galaxies with $20<m_{\rm AUTO}<24.5$ as a function of the separation to the nearest galaxy with $m_{\rm input}<26$ in the input catalogue (top) and selection bias (bottom) for a weight function with $\sigma_{\rm w}=2$ pixels (black) and $\sigma_{\rm w}=3$ pixels (grey). {\it Right panel}: idem, but now as a function of distance to the nearest detected galaxy. The insets in the panels zoom in on the results for separations larger than $2''$. The solid lines in the bottom panels show the corresponding direct measurements of the selection bias. 
\label{fig:metacal_sep}}
\end{figure*}

This suggests that \metacal\ cannot fully account for the shear bias that is introduced by blending. Also the clear difference between the observed and inferred selection bias for the baseline case suggests that this is not correctly estimated (the agreement is much better for the grid simulations, shown in red). To explore this further we  compute the post-\metacal\ bias 
as a function of separation to the nearest galaxy in the input catalogue ($m<26$) and show the results in the left panel of Fig.~\ref{fig:metacal_sep}.

Both choices for $\sigma_{\rm w}$ yield very similar results, except for very small values for $r_{\rm sep}$ where the larger weight function suffers more from blending, resulting in somewhat larger net biases. In both cases the bias rises quickly for separations $r_{\rm sep}<2''$  and becomes highly negative for $r_{\rm sep}<1''$, suggesting that it may be wise to exclude such galaxies from the cosmic shear analysis, if possible. As the bottom panel in Fig.~\ref{fig:detbias_separation} shows, this implies a 30\% reduction in the galaxy number density, so that one may want to allow for larger residual biases, although the gain may still be limited because undetected blends also tend to increase the shape noise \citep{Dawson16}.

The inset in the top panel shows that for $r_{\rm sep}>2''$ the bias is small: we find a mean bias $\langle\mu\rangle=0.000\,47\pm 0.000\,19$, whereas the bias for the full sample is $\langle\mu\rangle=0.002\,88\pm 0.000\,29$ (see Table~\ref{tab:metacal_bias} for more results). This confirms that \metacal\ can provide (nearly) unbiased shear estimates for isolated galaxies. Unfortunately, in practice we do not know whether or not a galaxy is blended, and the right panel of Fig.~\ref{fig:metacal_sep} shows the results for a more realistic scenario. 

The bias as a function of distance to the nearest detected galaxy shows a similar dependence for small separations as in the left panel, but the biases peak at larger values. Both weight functions yield consistent biases, even though the estimated selection biases differ (bottom panel). More importantly, for $r_{\rm sep}>2''$ the bias no longer vanishes. Many of the blends are not identified as such, resulting in a bias of $\langle\mu\rangle=-0.006\,49\pm 0.000\,22$ for apparently isolated galaxies. This is maybe not too surprising, because the \sextractor\ detection bias for apparently isolated galaxies 
(open grey points in Fig.~\ref{fig:detbias_separation}) did not converge to the value when galaxies are placed on a grid.

As this is perhaps the cleanest sample of sources that could be identified in a survey, our results imply that an algorithm that can provide unbiased shear estimates under ideal circumstances will still be significantly biased in reality. This also has implications for machine learning approaches \citep{Gruen10, Tewes19,  Pujol20}, which will have to be trained on simulations that include realistic blending.

Our findings suggest that, while \metacal\ is able to account for selection bias for isolated galaxies, blending limits the performance in more realistic scenarios. The image simulations can, however, be used to account for these residual biases, provided the simulations capture the complexities of real data. We therefore explore the sensitivity of the post-\metacal\ bias to changes in the simulation inputs, similar to what we did for the \sextractor\ detection bias.

\henk{The red points in Fig.~\ref{fig:nfac} show that the sensitivity to the galaxy density, captured by $n_{\rm fac}$, has changed sign compared to the \sextractor\ detection bias, but the amplitude of the trend is similar with $\partial\mu_{\rm meta}/\partial n_{\rm fac}=0.003\,75\pm 0.000\,33$, suggesting that it remains important to use the correct galaxy density in the simulations.} The changes in multiplicative bias as a function of $f_{\rm size}$ and $\epsilon_{\rm fac}$ are shown as red points in Fig.~\ref{fig:morphology}. Indeed we find that the sensitivities to these morphological parameters are reduced significantly compared to the \sextractor\ detection bias, with $\partial\mu_{\rm meta}/\partial f_{\rm size}=0.0087\pm0.0022$ and $\partial\mu_{\rm meta}/\partial \epsilon_{\rm fac}=0.0005\pm 0.0027$. Similarly we find no clear change in bias if we replace the \sersic\ index by a single value (red points in Fig.~\ref{fig:bias_sersic}). The sensitivity to the truncation of the surface brightness profile is enhanced, as indicated by the red points in Fig.~\ref{fig:bias_galtrunc}, but the bias converges for $r_{\rm trunc}>4$. 
\begin{figure*}
\centering
\leavevmode \hbox{%
  \includegraphics[width=8.5cm]{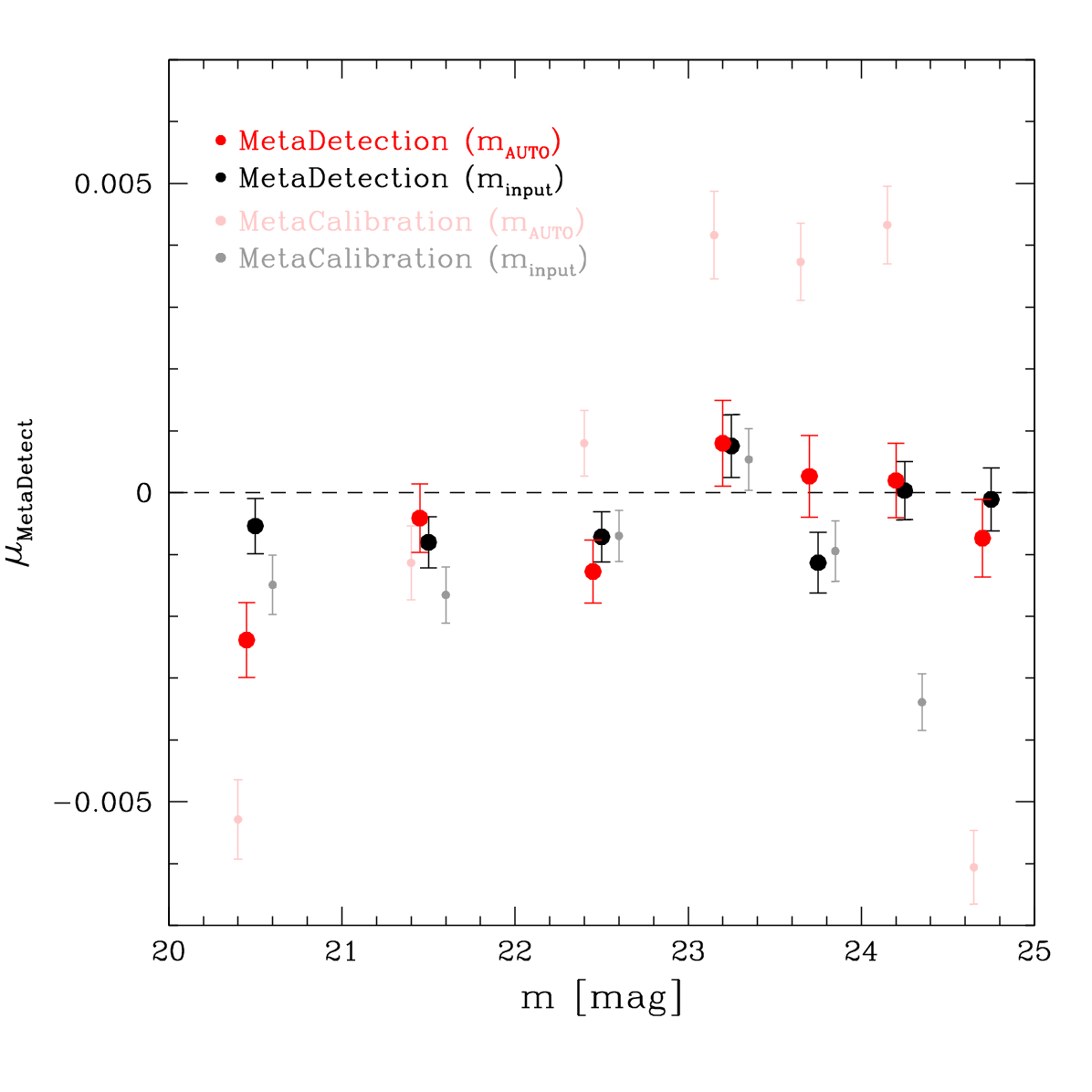}
   \includegraphics[width=8.5cm]{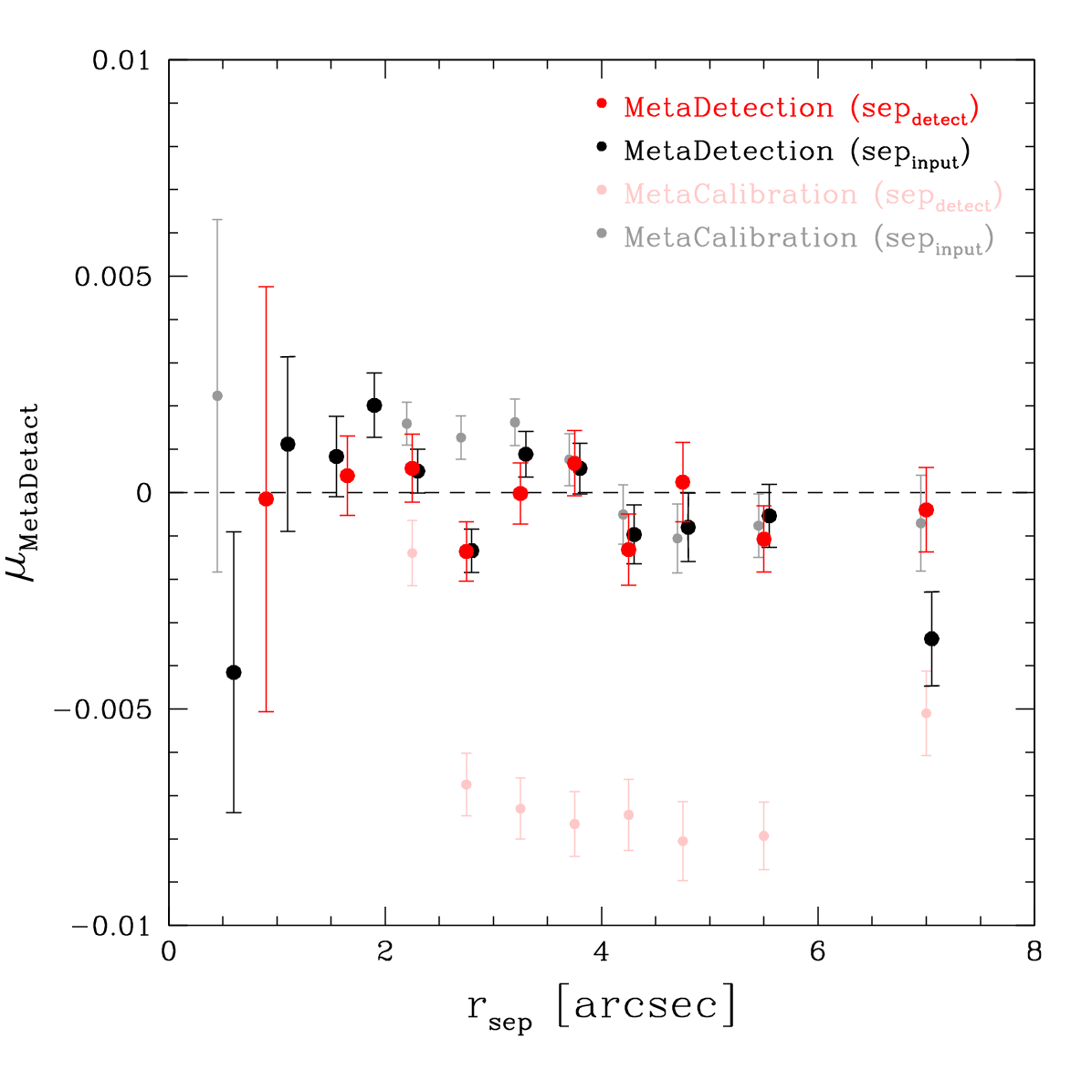}}
\caption{{\it Left panel}: multiplicative bias after \metadet\ as a function of magnitude, with galaxies selected by the input magnitude (black) or the observed magnitude (red). {\it Right panel}: multiplicative bias for galaxies with $20<m_{\rm AUTO}<24.5$ as a function or $r_{\rm sep}$, the distance to the nearest neighbour in the input catalogue (black points) and the distance to the nearest detected galaxy (red points). The light coloured points indicate the corresponding results for \metacal. In the case of \metadet\ the biases show no trend with magnitude or distance to the nearest galaxy, and are consistent with zero.
\label{fig:meta_detect}}
\end{figure*}

\section{MetaDetection}
\label{sec:meta_detect}

The results presented in Table~\ref{tab:metacal_bias} and Fig.~\ref{fig:metacal_sep} show that undetected blending is a significant source of bias, even for space-based Stage IV surveys. High quality, deep observations can help improve the fidelity of the image simulations that are used to quantify this residual bias, and our results indicate that the sensitivity to the simulation inputs are relatively small, but it would be better if this could be avoided in principle.

\cite{Sheldon19} proposed an alternative implementation of the \metacal\ approach where one effectively bypasses the steps to estimate $\bm{\mathsf{R}}^\gamma$ and $\bm{\mathsf{R}}^{\rm S}$. This approach, called \metadet, uses the same sheared images, but both the detection and the shape analysis are performed on these images. By avoiding the use of the unsheared image as a reference, the detection biases should vanish. The downside, however, is the lack of such a reference image, which complicates the labelling of galaxies that is needed to associate them with a tomographic redshift bin. 

We apply \metadet\ to our simulated images and find that the resulting average bias for galaxies with $20<m_{\rm AUTO}<24.5$ is very small: $\langle\mu\rangle=0.000\,01\pm 0.000\,30$ (we report the results for the individual shear components in Table~\ref{tab:metacal_bias}). 
\henk{We note that we have not quantified how this result changes when we shear the scene when creating the images (see Sect.~\ref{sec:estimates}), but the results presented in \cite{Sheldon19} suggest that this difference should be small for the much smaller \Euclid\ PSF.}

The left panel in Fig.~\ref{fig:meta_detect} shows the bias as a function of observed magnitude ($m_{\rm AUTO}$; red points) and input magnitude
($m_{\rm input}$; black points). The average biases are small and do not depend on magnitude, even for galaxies as faint as $m=25$. For reference, we indicate the corresponding \metacal\ results by the light coloured points. \henk{This is encouraging, because one could imagine 
estimating photometric redshifts for the galaxies in each of the five \metadet\ catalogues, which could subsequently be used to assign them to tomographic bins.} How to incorporate this into a full cosmic shear analysis is beyond the scope of this paper, but it is clearly worthwhile to explore further.

The potential of \metadet\ is confirmed further by the right panel of Fig.~\ref{fig:meta_detect}, where we show the bias as a function of distance to the nearest neighbour in the input catalogue (black) and the nearest detected neighbour (red). The improvement with respect to the \metacal\ case, indicated by the light red coloured points, is evident: \metadet\ is able to account for the blending of galaxies, resulting in residual biases that meet the stringent requirements for Stage IV surveys \citep{Cropper13}. Moreover, the blue points in Figs.~\ref{fig:nfac} and~\ref{fig:bias_galtrunc} show that the bias after \metadet\ no longer depends on the galaxy density $n_{\rm fac}$ or the truncation radius $r_{\rm trunc}$.

\section{Discussion}
\label{sec:discussion}

Our results show that the detection of galaxies results in a significant source of bias for weak lensing surveys \citep[also see][]{FC17,Kannawadi19,Hernandez20}. Although both survey characteristics and galaxy morphologies play a role, it is clear that undetected blending is the main concern for Stage IV surveys. In particular, we used \metacal\ as a proxy for a perfect shape measurement algorithm, and showed that this problem persists also in this case. Nonetheless, the reduced sensitivity to the simulation setup indicates that image simulations can provide accurate estimates of residual biases. Such simulations may be needed regardless, because \metacal\ cannot account for all sources of bias \citep{Huff17}. 

As discussed in \cite{Huff17} the image manipulation step assumes that the image is linearly related to the true surface brightness distribution. A wide range of instrumental effects limit the accuracy of this assumption. Some of these can be partially corrected for during the image processing, but the impact of their residuals should also be assessed using sufficiently realistic image simulations. A particular concern for \Euclid\ and the {\it Roman} Space Telescope is the fact that the pixel scale is relatively large compared to the PSF. This is not a problem \emph{per se} for the PSF itself, as a well-sampled model might be inferred from the data, but galaxies with small observed sizes might still be affected.
However, \cite{Kannawadi20}, with a larger fraction of small galaxies in their input catalogue, show that the bias due to undersampling is effectively mitigated when using a weight function with $\sigma_{\rm w} \ge 0\farcs{15}$. This is consistent with the absence of any significant residual biases in this work, and therefore, undersampling of small galaxies need not be a major concern for \Euclid.

In the case of \Euclid, charge-transfer inefficiency and the presence of cosmic rays also bias the shape measurements. Also blending/contamination by stars affects the shear bias \citep{Hoekstra17}, whereas spatial variations in the colours of galaxies lead to colour-gradient biases \citep[e.g.][]{Semboloni13, Er18}. These biases are also present after \metadet, which does provide unbiased shear estimates for our simulated \Euclid-like simulations. 

Our results suggest that \metacal\ and/or \metadet, combined with sufficiently realistic image simulations, provide a viable way forward towards accurate shear estimates for Stage IV surveys. Many practical complications remain, and we briefly review some of these here. We start by examining the computational needs: \Euclid\ aims to measure the shapes of over two billion galaxies, which places constraints on the time it takes to measure a galaxy shape. We apply the \metacal-step to the full images, and run the object detection algorithm on the  \metacal\ images, using the output for the unsheared image as our new detection catalogue. The computational needs are driven by the image manipulation steps,
which take about 150 seconds for each $4000\times 4000$ pixel image on a single core of an Intel Xeon Gold 5115 2.4 GHz CPU in a Dell R840 server (equipped with 80 cores). This included some I/O because we saved the five images to disk for \sextractor, which in principle can be avoided. The five \sextractor\ calls take on average 16 seconds and the shape measurements themselves take a total of 7 seconds for the baseline case. This amounts to a total processing time of about 0.06 second per galaxy on a single core.  In this paper we use \metacal\ to correct for the convolution with an isotropic PSF, but it can be extended to correct for PSF anisotropy (\citetalias{Sheldon17}). This requires 4 more images to be created, which approximately doubles the runtime of the \metacal-step. 

The memory needs of our current setup are substantial when creating the \metacal\ images, requiring about 18 Gb per core. This prevented us from using all available cores. In practice the analysis will have to be performed on much smaller postage stamps, because the PSF will vary across the field-of-view.
In fact, in the case of \Euclid, the PSF $P$ to use is the SED-weighted one.
Although the PSF in this case varies from object to object, it can be uniquely estimated from unresolved multi-band data \citep{Eriksen18}. As it is important that the effects of blending can be captured, the postage stamp should be at least be $8''\times 8''$. This estimate is based on the fact that the \metacal\ bias converges for $r_{\rm sep}>3''$ (see the right panel of Fig.~\ref{fig:metacal_sep}). \cite{Eriksen18} show that the effective \Euclid-PSF size varies by at most about 2\%, which suggests that using a single PSF for such a postage stamp would still capture the bias caused by blending. 

For our galaxy number density a postage stamp of $8''\times 8''$ means that the total number of pixels that needs to be manipulated increases by about 30\%. Given the reduced memory needs this would actually allow more cores to be used by a typical server. Including the correction for PSF anisotropy, we thus estimate that analysing $2\times 10^9$ galaxies would take about 70 days on our benchmark server with 80 cores. We note that this is a bare minimum, because one may want to analyse the individual exposures instead. Nonetheless, these estimates suggest that it may be possible to apply \metacal\ to Stage IV data sets. Alternatively, \metacal\ or \metadet\ can be applied to subsets of data to provide bias estimates for machine learning algorithms. Once trained these can estimate shapes very quickly \citep[e.g.][]{Pujol20}. 

Finally we note that \metacal\ and \metadet\ allow us to obtain unbiased shear estimates, but intrinsic alignments of galaxies
prevent a straightforward interpretation of the lensing signal 
\citep[see, e.g][for a review]{Joachimi15}. Direct observational constraints on the intrinsic alignment signal rely on accurate ellipticity measurements \citep{Georgiou19a}. Moreover the strength of the alignment signal depends on the shape measurement itself \citep[e.g.][]{Georgiou19b}. Hence care has to be taken when using physically motivated priors for the intrinsic alignment signal \citep{Johnston19, Fortuna20} in a cosmological analysis when the shear estimates are based on an intrinsically biased shape estimator, like the one we adopted here. 

\section{Conclusions}

Accurate measurements of the shapes of galaxies are a key ingredient for weak gravitational lensing studies. As a consequence improving the fidelity of the shape measurement algorithms has received much attention. Application of these algorithms to simulated data have played an important role in improving the performance. It has also become clear that it is important that the simulated data resemble the observations closely \citep[see e.g. the discussion in][]{Kannawadi19}. \citetalias{Hoekstra17} presented a detailed study for a simulated \Euclid-like data set, highlighting the challenges in ensuring sufficient realism. 

In this paper we use \Euclid-like image simulations, similar to the ones studied in \citetalias{Hoekstra17} to examine another important source of bias, which is present even if the shapes estimates are perfect. Detection bias arises because
the probability with which an object is detected (or selected) in an image depends on the shear. This has been known for quite a while \citep[e.g.][]{Hirata03}, but its contribution to shear bias has been largely ignored until recently. We find that the bias is generally smaller than instrumental bias, but it does lead to  multiplicative biases in the shear that exceed requirements for the next generation of cosmic shear surveys. 

To quantify the size of the bias we used \sextractor\ \citep{Bertin96} to detect objects. We matched the resulting catalogues to the input catalogue from which we took the true ellipticities. This mimics the performance of an ideal shape measurement algorithm. As reported in Table~\ref{tab:bias} we find that the average shear is underestimated by about 1\%; five times larger than can be tolerated for Stage IV surveys \citep{Cropper13}. This result is robust against changes in the settings of the detection algorithm. A smaller detection bias, which only affects the faintest galaxies, is observed when we place galaxies on a grid. This is caused because galaxies that are oriented perpendicular to the shear are detected preferentially. In the case of an anisotropic PSF, we find a small positive additive detection bias because galaxies that align with the PSF are detected with a higher significance.

The larger bias in our baseline simulation, where galaxies are placed randomly, is caused by the blending of sources, with biases exceeding 2\% for separations less than $1''$. Deeper observations can reduce the detection bias, but blending introduces a floor that still exceeds requirements. Following \citetalias{Hoekstra17} we also explored how the detection bias depends on the simulation inputs. We find that the detection bias increases linearly with galaxy density, the result of the higher occurrence of blending. The bias is also reduced when the galaxies are smaller or rounder. We observe a slight dependence of the surface brightness profile (quantified by the \sersic-index $n$). It is, however, important that the galaxy profiles are not artificially truncated before 4 effective radii.

Although the detection bias is far less sensitive to variations in the simulation parameters compared to the KSB algorithm studied in \citetalias{Hoekstra17}, the realism of the simulations, in particular ensuring that the variety in galaxy morphologies is adequately captured, remains a concern. We therefore explored the performance of an alternative approach that uses the data to determine the response of an ensemble of galaxies to a shear. This so-called \metacal\ was recently developed by \cite{Huff17} and \citetalias{Sheldon17} and showed promise for isolated galaxies. The problem of blending was investigated in more detail by \cite{Sheldon19} who showed that a variation of \metacal, dubbed \metadet, can be used to address this. 

We found that \metacal\ provides a (near) perfect shear estimate in the absence of detection bias. Importantly, the choice of shape measurement algorithm is irrelevant and we opt for weighted quadrupole moments with a fixed width for the Gaussian weight function. For isolated galaxies the performance of \metacal\ is only limited by the accuracy of the PSF model (which we assume to be perfect) and biases introduced by the pixelisation of the images (which are also negligible in our case). For the grid-based images we obtained a mean multiplicative bias of $\langle\mu\rangle=0.000\,64\pm 0.000\, 29$, well within requirements for Stage IV surveys. However, blending will limit the actual performance and for our baseline case we measured a significant bias of $\langle\mu\rangle=0.002\,88\pm 0.000\, 29$. We showed that this is caused by blended objects, many of which cannot be identified as such. In fact selecting galaxies that appear isolated (no detected neighbour within $2''$) leads to a larger net bias of $\langle\mu\rangle=-0.006\,49\pm 0.000\, 22$.
The post-\metacal\ bias is less sensitive to changes in the input galaxy sizes or the ellipticity distribution, but does still depend on the galaxy number density. Nonetheless, these findings suggest that image simulations can be used to account for the residual biases in \metacal. Such simulations are needed anyway to determine the biases caused by  instrumental effects. Moreover, simulations will be essential to understand the correlation between shear bias and biases in photometric redshifts that blending should introduce.

{\sc MetaDetection} uses the same sheared images as \metacal, but both the detection and shape analysis are performed on these images. The resulting multiplicative bias for galaxies with $20<m_{\rm AUTO}<24.5$ is very small: $\langle\mu\rangle=0.000\,01\pm 0.000\,30$. Moreover, the bias does not depend on magnitude or distance to the nearest neighbour, indicating the blending does not bias the mean shear. The lack of a reference catalogue, which otherwise would re-introduce the selection bias, may lead to practical complications. However, it may be possible to assign photometric redshifts to the different \metadet\ catalogues and define tomographic redshifts for each catalogue. Alternatively, the \metadet\ estimates for various selections of source can act as reference values for machine learning approaches.
More work is needed to examine the practical implementation of both \metacal\ and \metadet, but our results suggest that these, combined with sufficiently realistic image simulations, provide a viable way forward towards accurate shear estimates for Stage IV surveys.

\vspace{0.5cm} 

The authors are grateful to Erin Sheldon and the developers of \galsim\ for making their software packages publicly available. \henk{They also thank Tim Schrabback and an anonymous referee for useful comments that helped improve the paper.} HH and AK acknowledge support from  the Netherlands Organisation for Scientific Research (NWO) through grant 639.043.512. HH and TDK acknowledge support from the EU Horizon 2020 research and innovation programme under grant agreement 776247.

\bibliographystyle{aa}
\bibliography{detect}

\appendix

\section{Sensitivity to detection setup}
\label{app:setup}

\begin{figure*}
\centering
\leavevmode \hbox{%
  \includegraphics[width=18cm]{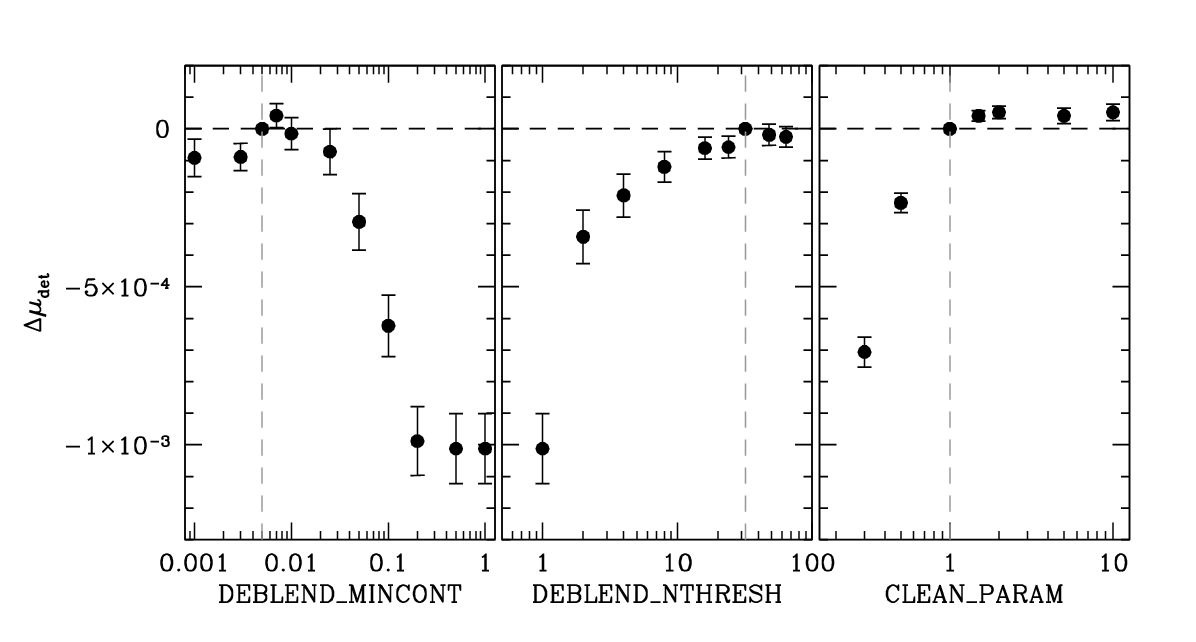}}
\caption{Change in multiplicative shear bias $\Delta\mu$ as a function of the \sextractor\ parameters that affect the deblending of objects. The vertical grey dashed line indicates the baseline value (also see Table~\ref{tab:setup}).
These default values result in detection biases that are close to optimal. \label{fig:settings}}
\end{figure*}

Several parameters influence the deblending of objects by \sextractor, and we examine their impact on the detection bias here. Compared to the choice of filter function
in the detection step (see Fig.~\ref{fig:detbias_filter}), the changes in bias
are smaller, but as Fig.~\ref{fig:settings} shows, they can still change by as much as $10^{-3}$ in the most extreme cases. The changes are negligible, however, when the parameters remain close to their default values. 

As described in detail in \cite{Bertin96} \sextractor\ uses multi-thresholding to separate objects that were extracted as a single object during the detection step. The pixels that make up an extracted object are thresholded by {\tt DEBLEND\_NTHRESH} levels that are spaced exponentially between the extraction threshold and the peak value; a low value reduces the effectiveness of  the deblending step. A tree model of the surface brightness is created \citep[see Fig.~2 in][]{Bertin96} and the model works its way down to the trunk, deciding at each junction whether or not to split the object into separate ones. This decision is governed by the value of {\tt DEBLEND\_MINCONT}, which is the minimum fraction of the flux that needs to be contained in the deblended source; hence a high value of this parameter means that only sources of similar brightness are deblended. 

The left and middle panels in Fig.~\ref{fig:settings} show the change in $\mu_{\rm det}$ when we use different values for
{\tt DEBLEND\_MINCONT} and {\tt DEBLEND\_NTHRESH}, respectively. We see that the bias increases by about 
$10^{-3}$ if the deblending is minimised. The biases barely change if we vary the parameters about the baseline settings (indicated by the vertical grey dashed lines). 

Noise in the images may result in the outer regions of sources to be broken up into smaller pieces. Such inadverted `deblending' is undone by cleaning the catalogue. For each object, \sextractor\ estimates the contribution from neighbouring galaxies to the mean surface brightness assuming a Gaussian extrapolation of their profile, 
and subtracts this from the object in question. If it is still above the detection threshold, the object is accepted. 
The width of Gaussian used to extrapolate the flux from nearby galaxies can be changed from its default estimate by
{\tt CLEAN\_PARAM}. The right panel in Fig.~\ref{fig:settings} shows that that little cleaning (values less than 1) rapidly increases the detection bias, whereas more aggressive cleaning has little impact. 

\section{Relation between additive and multiplicative bias}
\label{app:additive}

If one is concerned about a particular instrumental effect that might introduce additive bias, one can simply average the shear estimates in the appropriate coordinate system (e.g. the one defined by the detector), because the cosmological signal should vanish if enough data are included. For instance, \cite{Hoekstra11b} used this approach to remove the 
additive bias caused by charge transfer inefficiency  (CTI) in  {\it Hubble} Space Telescope observations, and \cite{Hildebrandt20} use this to account for an additive bias that arises from the shape measurement algorithm
\citep[as shown in][]{Kannawadi19}. However, this empirical approach ignores the fact that such systematics may cause multiplicative bias as well, as we show here.

To do so we express the  observed shape of an object in terms of the unweighted quadrupole moments $Q_{ij}$ of its surface brightness distribution \citep[e.g.][]{Massey13}. These can be combined into the polarisation, which has two components $\chi_i$ defined as:
\begin{equation}
\chi_1=\frac{Q_{11}-Q_{22}}{Q_{11}+Q_{22}},~\rm{and}~\chi_2=\frac{2Q_{12}}{Q_{11}+Q_{22}},
\end{equation}

If we now consider a (residual) effect that changes the observed quadrupole $Q'_{11}=Q_{11}+\delta Q_{11}$, while leaving the other moments unchanged\footnote{For the purpose of this derivation we are free to choose a convenient coordinate system.}, the observed polarisation is:
\begin{equation}
\langle\chi_1^{\rm obs}\rangle\approx
\langle\chi_1^{\rm true}\rangle\left({1-\frac{\delta Q_{11}}{Q_{11}+Q_{22}}}\right)+\frac{\delta Q_{11}}{Q_{11}+Q_{22}},
\label{eq:cbias}
\end{equation}
and
\begin{equation}
\langle\chi_2^{\rm obs}\rangle\approx
\langle\chi_2^{\rm true}\rangle\left({1-\frac{\delta Q_{11}}{Q_{11}+Q_{22}}}\right).
\end{equation}

The last term \henk{in Eq.~(\ref{eq:cbias})} corresponds to the additive bias $c_1$, whereas \emph{both} polarisation components are biased low by a factor $(1+\mu)$. In this simple case we find that $\mu_1=\mu_2=\mu=-c_1$. Hence, instrumental effects that introduce additive shear bias by modifying the recorded images generally also cause a multiplicative bias that is similar in amplitude, affects both shear components, but has the opposite sign. 

\begin{figure}
\centering
\leavevmode \hbox{%
  \includegraphics[width=8.5cm]{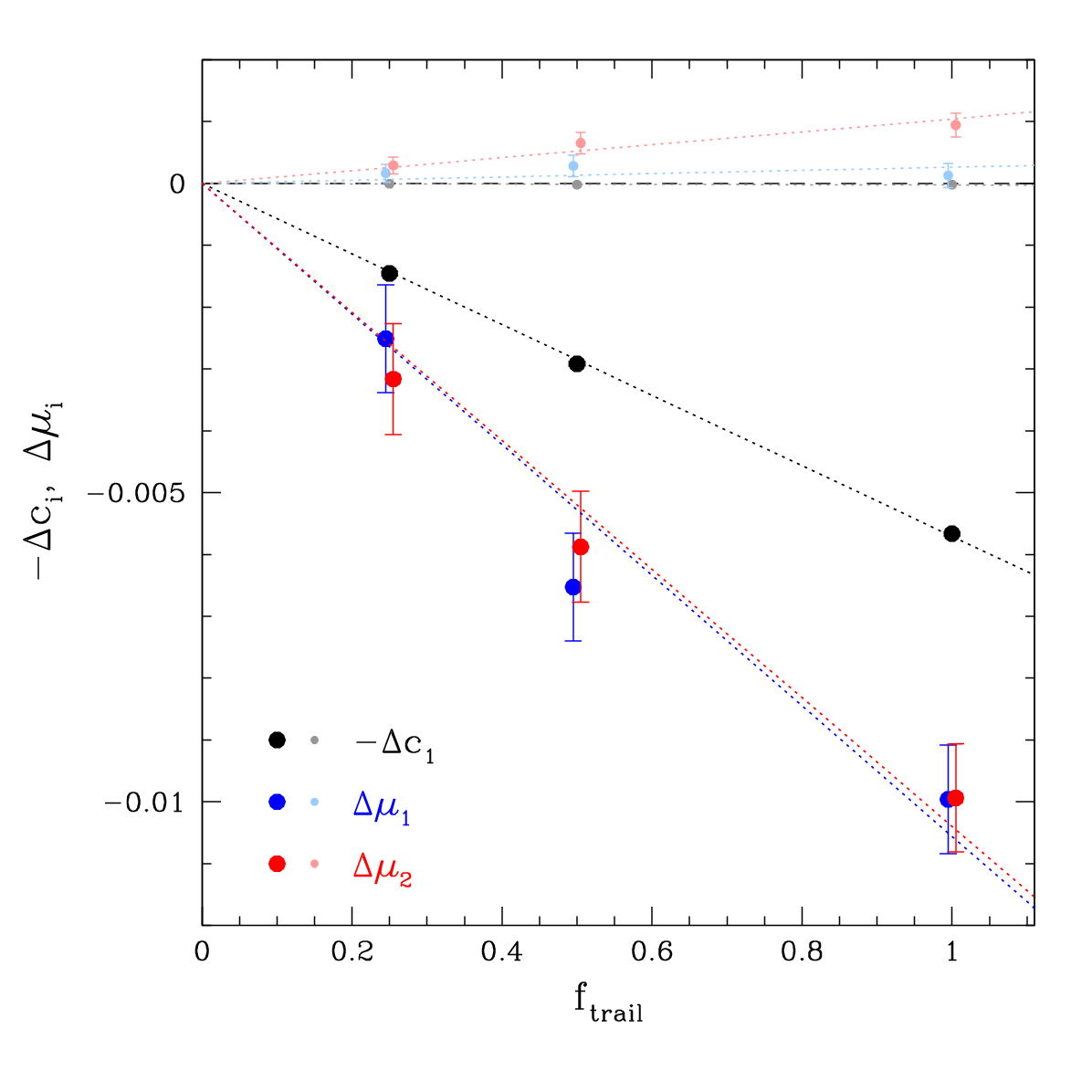}}
\caption{Comparison of the change in multiplicative bias and additive bias when some of the charge is trailed during the readout process. The amount of trailing is determined by the value of $f_{\rm trail}$. The light coloured points show the (small) detection biases, whereas the bright coloured points show the 
$-\Delta c_1$ (black), $\Delta\mu_1$ (blue) and $\Delta\mu_2$ (red). The amplitude of the additive bias is about half of the multiplicative bias, but has the opposite sign, as predicted.
  \label{fig:bias_trail}}
\end{figure}

To verify this result, we created images where we mimic the effect of charge trailing, which in reality might be caused by dielectric absorption in the read-out electronics \citep{Toyozumi05}. Rather than computing the actual change in bias voltage, we simply assume that the amount of charge
that is added to the next pixel in the $i$th column is given by a power law, so that 
$$f(i,j)=f(i,j)+f_{\rm trail}\, f(i-1,j)^{0.4},$$
where the value for the power law slope is inspired by what is observed in OmegaCAM data (Hoekstra et al., in prep.), and $f_{\rm trail}$ is the amplitude. To create the images we add a realistic background level, compute the trailed image, add this to the original image and finally subtract the background again. We analyse the resulting images as before. 

Figure~\ref{fig:bias_trail} shows the resulting additive and multiplicative bias as a function of $f_{\rm trail}$, where we note that the applied values are unrealistically large to ensure a signal that could be measured using \metacal. The light coloured points show the detection biases that we observe, which are negligible, with the exception of $\mu_2^{\rm det}$. The trend of $\mu^{\rm det}_2$ with $f_{\rm trail}$ is largest if we consider only isolated galaxies.
The trailing changes both the centroid and the flux of the galaxy, both of which will change the multiplicative bias somewhat, but it is not obvious why this does not affect $\mu_1^{\rm det}$. Nonetheless the detection biases are small, even for this extreme level of charge trailing. 

More relevant for the discussion here are the bright coloured points, which show the biases after \metacal. As predicted, the changes in
$\mu_1$ (blue) and $\mu_2$ (red) are consistent, and the sign of $\Delta c_1$ is opposite from $\Delta\mu$. The change is about half of what is predicted by Eq.~(\ref{eq:cbias}),but we note that it is no longer applicable when weighted moments are used to measure the shapes, and corrections of $\mathcal{O}(1)$ are expected. In fact, tests with an elliptical Airy PSF, for which the unweighted quadrupole moments do not converge, show that the additive PSF biases are smaller by as much as a factor 4. This lower sensitivity to PSF anisotropy implies that allocating a residual bias of $|c_{\rm PSF}|<1.5\times 10^{-4}$
corresponds to a tolerable error in the PSF ellipticity of $|\Delta\epsilon_{\rm PSF}|<5.8\times 10^{-4}$ instead of $<2\times 10^{-4}$ adopted by \cite{Cropper13}, \henk{but we caution that the sensitivity depends on the PSF profile. For instance when we used PSF models that included various aberrations, we typically found larger residual biases. In all cases, however, the sensitivity was lower compared to the estimate based on unweighted moments.}

An empirical correction for additive bias should therefore be considered with these limitations in mind. Although our results for this particular case show that the multiplicative bias is still within a factor 2 of our naive prediction, it cannot replace a proper physical modelling and calibration of instrumental effects. One important reason to understand the cause of any residual additive bias is that it may not be cleanly separable from other biases; this would complicate estimating the impact on the multiplicative bias. Nonetheless, provided that the biases are small to begin with, we expect that the multiplicative biases will be similar in amplitude to the additive bias. This is still helpful, because  \cite{Kitching19} showed that the impact of such small scale-dependent multiplicative biases is reduced further when we consider the power spectrum estimates used in cosmological analyses. 

\section{Uncertainty in input galaxy number density}
\label{app:galden}

\begin{figure}
\centering
\leavevmode \hbox{%
  \includegraphics[width=8.5cm]{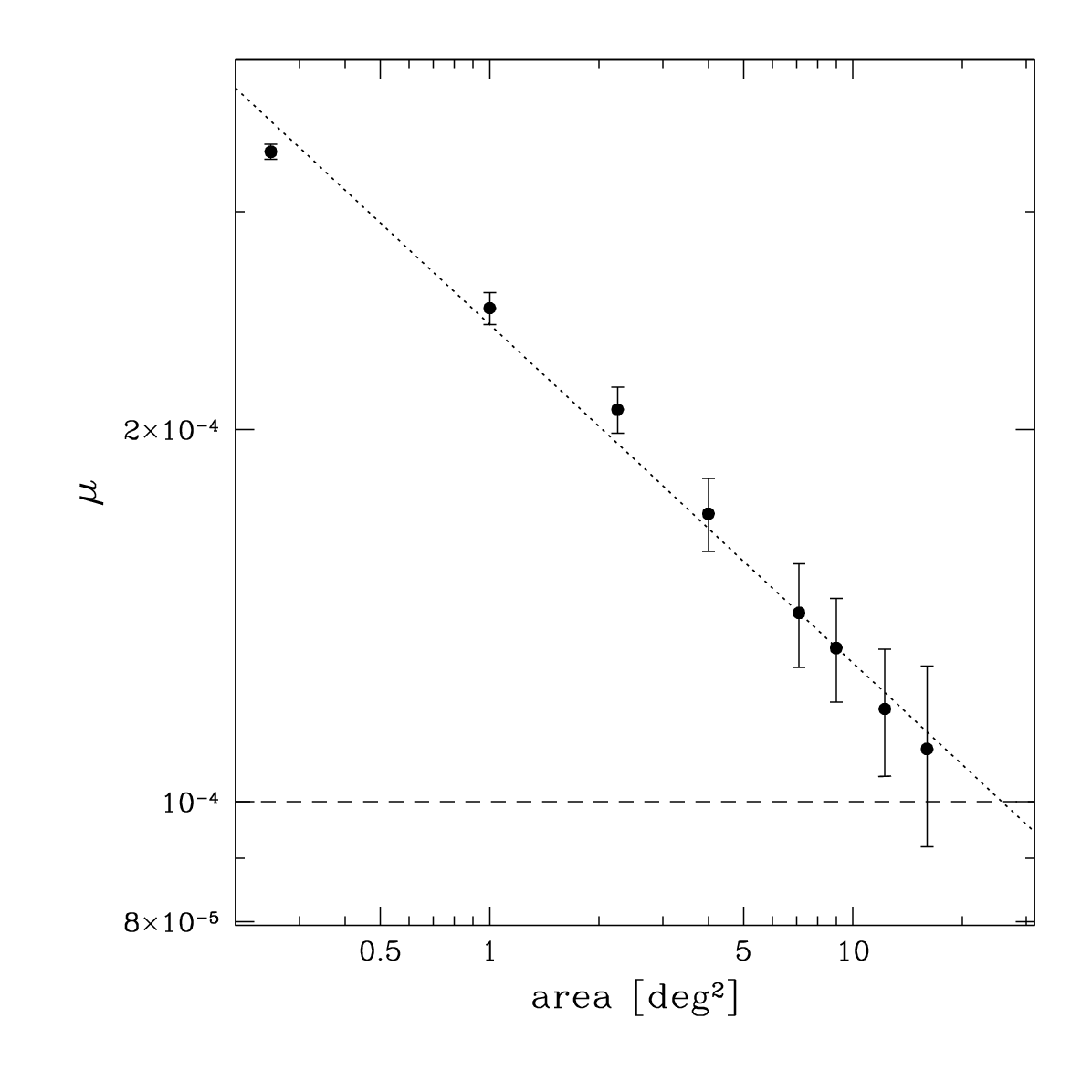}}
\caption{Multiplicative bias that arises from uncertainties in the average galaxy number density, as a function from the area used to determine the average density, assuming the sensitivity of $\mu$ with $n_{\rm fac}$ after \metacal\ (dashed red line in Fig.~\ref{fig:nfac}).
  \label{fig:mu_area}}
\end{figure}

As shown in Fig.~\ref{fig:nfac} the multiplicative bias depends on the number density of galaxies in the simulated images. Moreover, the results indicate that \metacal\ is unable to fully remove such a dependency, but in fact introduces a
weak positive dependence as $\partial\mu_{\rm meta}/\partial n_{\rm fac}=(0.0039\pm0.0005)$. Here we examine
what area needs to be observed so that the uncertainty in the observed value of $n_{\rm fac}$ leads to a bias in the multiplicative bias of $|\delta\mu|<10^{-4}$.

To estimate the expected variation in galaxy density as a function of angular scale we use the second data release
of the Marenostrum Institut de Ci{\`e}ncies de l'Espai (MICE)  grand challenge galaxy and halo light-cone simulation.\footnote{MICECATv2 is publicly available at \url{https://cosmohub.pic.es/home}} The mock galaxy catalogue is obtained from a large N-body simulation, from which a light-cone is constructed  \citep[see][for details]{Fosalba15}. The simulation is populated with galaxies using a hybrid halo occupation distribution and abundance matching technique described in \cite{Crocce15} and \cite{Carretero15}.

The second data release includes a mock galaxy catalogue that is complete for current stage III surveys ($m_i<24$),
but restricted to $z<1.4$,  resulting in an average number density of about 26 galaxies arcmin$^{-2}$ brighter than $m_{\rm VIS}=24.5$ in the \Euclid-VIS band. Although the catalogue thus lacks high redshift galaxies, it is sufficient for our purposes because the spatial variations are larger at lower redshifts where a fixed angular scale probes a smaller  volume. We retrieved 9 patches, each $10\times 10$ degrees, to determine the dispersion in galaxy counts when we subdivide these data into smaller areas. The relative variation is a direct estimate for the dispersion in $n_{\rm fac}$,
which in turn can be converted into an estimate of the uncertainty in the multiplicative bias. 

If we wish that the contribution to the uncertainty in the multiplicative bias due to the uncertainty in the mean galaxy density is $<10^{-4}$, the observed sensitivity of the bias after \metacal\ implies that we need to know $n_{\rm fac}$ with a relative precision of about 2.6\%. If we consider the variation in galaxy counts in a patch of 1 deg$^2$ the MICE simulations yield a dispersion of 0.064, which agrees remarkably well with observed estimates of the variation in galaxy counts by \cite{Herbonnet19} on similar angular scales and depths. In contrast, the approximately 0.25 deg$^2$  covered by GEMS \citep{Rix04} would introduce on average a multiplicative bias of about $3\times 10^{-4}$, taking up a significant part of the overall budget specified in \cite{Cropper13}.

Figure~\ref{fig:mu_area} shows how the multiplicative bias $\mu$ depends on the observed area of sky used to estimate the mean density of galaxies in the image simulations. The error bars correspond to the dispersion in the measured counts in the patches. The results are well described by a power law with a slope of  $-0.27$; the fit indicates that to achieve a bias $<10^{-4}$ we need to measure the galaxy counts in an area of about 30 deg$^2$. \henk{This estimate may be somewhat optimistic because we did not consider the impact of small clustering in our image simulations.} Of course the actual survey data can be used to validate the realism of the simulation, but in practice deeper observations of smaller areas are more useful as input to the image simulations. Hence, once can interpret these results as the minimum area for which deeper observations help to improve the fidelity of the image simulations.

\end{document}